\documentclass{ectj}

\usepackage{amsfonts,amssymb,graphics,epsfig,verbatim,bm,latexsym,amsmath,url,amsbsy,dsfont}
\usepackage{subcaption}
\usepackage{booktabs}
\usepackage{array}
\usepackage {tabu}
\usepackage{graphicx}
\usepackage{placeins}

\newtheorem{theorem}{Theorem}
\newtheorem{assumption}{Assumption}
\newtheorem{proposition}{Proposition}

\newtheorem{lemma}{Lemma}

\renewcommand{\thesection}{\arabic{section}}
\renewcommand{\theequation}{\arabic{section}.\arabic{equation}}

  \received{}
\title[Graphical Models]{Estimation of Graphical Models using the $L_{1,2}$ Norm}

\author[Chiong and Moon]{Khai X. Chiong$^{\dagger}$ and
                        Hyungsik Roger Moon$^{\ddagger}$}

\address{$^{\dagger}$Naveen Jindal School of Management, University of Texas at Dallas}
\email{khai.chiong@utdallas.edu}

\address{$^{\ddagger}$Department of Economics, University of Southern California \\
and School of Economics, Yonsei University}
\email{moonr@usc.edu}

\def\AmSTeX{$\cal A$\kern-.1667em\lower.5ex\hbox{$\cal M$}\kern-.125em
            $\cal S$-\TeX}
\def\BibTeX{{\rm B\kern-.05em{\sc i\kern-.025em b}\kern-.08em
            T\kern-.1667em\lower.7ex\hbox{E}\kern-.125emX}}

\graphicspath{{../Figures/}}

\newcommand{\argmax}{\operatornamewithlimits{argmax}}
\newcommand{\argmin}{\operatornamewithlimits{argmin}}

\newcommand{\Cov}{\mathrm{Cov}}

\newcommand{\tr}{\mathrm{tr}}

\newcommand{\pr}{\operatorname{Pr}}

\begin{document}

  \begin{abstract}
Gaussian graphical models are recently used in economics to obtain networks of dependence among agents. A widely-used estimator is the Graphical Lasso (GLASSO), which amounts to a maximum likelihood estimation regularized using the $L_{1,1}$ matrix norm on the precision matrix $\Omega$. The $L_{1,1}$ norm is a lasso penalty that controls for sparsity, or the number of zeros in $\Omega$. We propose a new estimator called {\em Structured Graphical Lasso} (SGLASSO) that uses the $L_{1,2}$ mixed norm. The use of the $L_{1,2}$ penalty controls for the {\em structure} of the sparsity in $\Omega$. We show that {\color{black} when the network size is fixed,} SGLASSO is asymptotically equivalent to an infeasible GLASSO problem which prioritizes the sparsity-recovery of high-degree nodes.  Monte Carlo simulation shows that SGLASSO outperforms GLASSO in terms of estimating the overall precision matrix and in terms of estimating the structure of the graphical model. In an empirical {\color{black} illustration} using a classic firms' investment dataset, we obtain a network of firms' dependence that exhibits the core-periphery structure, with General Motors, General Electric and U.S. Steel forming the core group of firms.
\\



  \end{abstract}

\section{Introduction}

{\color{black}
The Gaussian graphical model is a graph summarizing the conditional independence relationships among a group of random variables. Suppose that $(X_{1},\dots,X_{p})$ is distributed with $\mathcal{N}(\bm{0},\Sigma_{0})$. If the $(i,j)$-th entry of the precision matrix $\Sigma^{-1}_{0} \equiv \Omega_{0}$ is zero, i.e., $\Omega_{0,ij}=0$, then it is known that $X_{i}$ and $X_j$ are independent conditional on all other $X_{k}$, $k\neq i,j$ (e.g., Chapter 9 of \cite{Hastie2015statistical}). The Gaussian graphical model is then obtained by identifying the zeros of the precision matrix, and letting nodes $i$ and $j$ be linked if and only if the $(i,j)$-th entry of $\Omega$ is non-zero. 


In our empirical illustration, the variables $(X_{1},\dots,X_{p})$ represent the annual investment or the residuals of investment equations of the $p$ number of firms: $(1,\dots,p)$. If two firms $i$ and $j$ are {\em not} linked in this graphical model, it means that the investment decisions of firms $i$ and $j$ are independent of each other, conditional on all other firms $X_{k}$, $k\neq i,j$. Hence the investment decisions of firms $i$ and $j$ do not directly affect  each other. On the other hand, if two firms $i$ and $j$ are linked in the graphical model, then the investments of firms $i$ and $j$ have {\em direct} effects on each other, without the mediation of other firms.
In recent applications in economics, for instance, \cite{giudici2016graphical} uses graphical modeling to obtain the network of international financial flows. 



It is well known that the inverse of the standard sample covariance estimator, $\hat{\Sigma}^{-1},$ performs poorly particularly in recovering the underlying graphical structure. The reason is that the sample precision matrix produces a dense matrix without any zeros, and hence the resulting graphical model is always a complete network where all nodes are linked to all other nodes.

A widely-used technique to estimate graphical models is called Graphical Lasso (GLASSO), which aims to estimate the precision matrix $\Omega$ while imposing a sparsity constraint on $\Omega$. For an overview of this topic, see  \cite{Hastie2015statistical,cai2016estimating,fan2016overview}. 
In this paper, we propose an estimation method for graphical models, which works by estimating the precision matrix, $\Omega_0$, taking into account the {\em structure} of sparsity in $\Omega_0$. }
 
The standard GLASSO approach estimates the precision matrices while controlling for the sparsity of $\Omega_0$, i.e. the number of zeros in $\Omega_0$. However it does not consider when these zeros could be distributed in a certain pattern or structure in $\Omega_0$.  For instance, in many economic applications it is reasonable to think that in the graphical model, some nodes have very few links, while some hub-like nodes have many links to other nodes. Such settings are common in economic applications: for instance, we commonly observe the so-called small-world properties in social and economic networks (\cite{jackson2008social}).


 The main feature of our estimation method is to use the $L_{1,2}$-norm penalty, which corresponds to the second moment of the weighted degree of nodes, i.e. $\sum_{k=1}^{p} |\Omega_{ik}|$, for $i=1,\dots, p$. Notice that Lasso uses the $L_{1,1}$-norm penalty, which is proportional to the first moment of the weighted degree distribution. We derive the asymptotic distribution of the new estimator, called Structured Graphical Lasso (SGLASSO) when the sample size $T \rightarrow \infty$ and the dimension of $\Omega_0$, $p$, is fixed.

 The main theoretical finding is that {\color{black}when the network size $p$ is fixed}, SGLASSO is asymptotically equivalent to an infeasible GLASSO problem which penalizes an entry of $\Omega$ proportionally according to the influence of the entry, where the influence of an $(i,j)$-th entry is defined as the sum of the true weighted degrees of nodes $i$ and $j$ in the graphical model. Specifically, SGLASSO is asymptotically equivalent to a modified GLASSO where the element-by-element penalty factor for the $(i,j)$-th entry of $\Omega$ is proportional to $(d_{i} + d_{j})$, where $d_{i} = \sum_{k=1}^{p} |\Omega_{0,ik}|$ is the {\em true} weighted degree of node $i$. Note that $d_{i}$ is unknown to us, and hence it is infeasible to be implemented as a GLASSO estimator.
 
 {\color{black}By means of Monte Carlo simulations, we demonstrate} that SGLASSO performs better than GLASSO in finite samples. In terms of estimating the overall precision matrix, SGLASSO achieves lower Kullback-Leibler and Frobenius losses. Moreover, in terms of recovering the structure of the graphical model, we also show that SGLASSO outperforms GLASSO.

 Overall, we should use the $L_{1,2}$ norm in cases where we are more interested in the relationships among influential nodes. For instance, in a financial network that exhibits a core-periphery structure, we are more interested in the relationships between the major core banks, and less interested in the relationships between the minor peripheral banks. Knowing how shocks cascades and propagates among influential banks is key to understanding financial contagion (\cite{elliott2014financial}).

However if we were to remain completely agnostic about the empirical context, there are still reasons for using the $L_{1,2}$ norm. Here, the virtue of the $L_{1,2}$ norm is that it allows us to express lasso penalties that scale automatically according to degree influences (this insight is due to our main theoretical result). In practice, when we use the $L_{1,2}$ norm in conjunction with the $L_{1,1}$ norm, we are allowing for lasso penalties that have a constant factor and an increasing factor over the nodes' degrees. This can only improve fit. The analogy is that the $L_{1,1}$ norm is a linear first-order lasso penalty, and the $L_{1,2}$ norm is a quadratic lasso penalty. Moreover, our simulation shows that using just the $L_{1,2}$ norm can achieve significant improvements over the $L_{1,1}$ norm. This suggests that it is important to allow for lasso penalties that scale according to nodes' degrees, and not restricted to a constant lasso penalty.

The paper is organized as follows. Section 2 introduces the model, the $L_{1,2}$-norm penalty, and a motivating example. Section 3 presents the theoretical results. {\color{black}Section 4 provides a brief summary of the Monte Carlo simulation results. The full simulation results are available in the Supplemental Information. Section 5 contains an empirical illustration.} The Appendix contains technical proofs and derivations. 

  \subsection{Related literature}
  
 {\color{black}This paper is related to literature of estimating high-dimensional inverse covariance matrices. In addition to the aforementioned surveys, one can refer to \cite{banerjee2008model,friedman2008sparse,yuan2007model,rothman2008sparse,ravikumar2011high}.}
 
  A different class of procedure frames the estimation of graphical models as nodewise or pairwise regressions, where the Lasso or Dantzig selector can then be used to achieve variable selection and high-dimensional regularization. The relevant papers belonging to this class are \cite{cai2011constrained,cai2016estimating,meinshausen2006high,peng2012partial,ren2015asymptotic}. 
For ultra high-dimensional Gaussian graphical models, a recent approach is the  Innovated Scalable Efficient Estimation proposed by \cite{fan2016innovated}.
%

\cite{lam2009sparsistency} study a general class of estimators that nests the GLASSO. In particular, they consider $\argmax_{\Omega\succ0} \big\{\log{\det \Omega} - \tr(\Omega \hat{\Sigma}) - \sum_{i \neq j} p_{\lambda}(|\Omega_{ij}|)\big\}$, where $p_{\lambda}(\cdot)$ is a penalty function that depends on a regularization parameter $\lambda$. The SGLASSO estimator does not belong to this class because our penalty term $\|\Omega\|_{1,2}^{2}$ is not additively separable in the entries of $\Omega$. Hence, SGLASSO cannot be analyzed using the framework proposed in \cite{lam2009sparsistency}.\footnote{It is possible to incorporate other penalty functions into our estimator. Consider:
$\hat{\Omega}_{\text{sc-glasso}} :=  \argmax_{\Omega\succ0} \big\{\log{\det \Omega} - \tr(\Omega \hat{\Sigma}) - \|p_\lambda(\Omega)\|^{2}_{1,2} \big\}$, where $p_\lambda(\Omega)$ denotes the matrix (of the same dimension as $\Omega$) such that $p_\lambda(\Omega)_{ij} = p_\lambda(\Omega_{ij})$. For instance, we can then let $p_\lambda(\cdot)$ to be the Smoothly Clipped Absolute Deviation (SCAD) penalty function (\cite{fan2001variable}).}
%



 Another type of mixed norm that is widely used in the high-dimensional linear regression setting is the $L_{2,1}$-norm, also known as the Group Lasso (\cite{friedman2010note,yuan2006model}), which is useful when parameters are naturally partitioned into disjoint groups, and we would like to achieve sparsity with respect to whole groups. In contrast, our proposed $L_{1,2}$-norm is distinct from the $L_{2,1}$-norm, and has not been studied in the context of covariance estimation.

 \section{Setup}

Let $\bm{X} = (X_{1},\dots,X_{p}) \in \mathbb{R}^{p}$ be a $p$-dimensional vector distributed according to a multivariate distribution with mean zero and covariance matrix $\Sigma_{0}$. Our goal is to obtain the graphical model corresponding to $\bm{X}$, which is defined as follows. 

The (Gaussian) graphical model of $\bm{X}$ is an undirected graph $G = (V,E)$. The set of nodes (or vertices) of this graph is $V = \{1,\dots,p\}$. Each node $i$ corresponds to a random variable $X_{i}$. The set of links (or edges) $E$ is such that $\{i,j\} \notin E$ if and only if $X_{i} \perp X_{j}$ conditional on all $X_{k}, k \in V \setminus \{i,j\}$. That is, the graphical model is defined such that there is {\em no} link between nodes $i$ and $j$ in the graph if and only if $X_{i}$ and $X_j$ are independent conditional on all other nodes besides $i$ and $j$. As such the graphical model $G$ summarizes the pairwise conditional independence relationships among $(X_{1},\dots,X_{p})$. The existence of a link in $G$ amounts to conditional dependence between the two nodes given all other ones.

Estimation of graphical models is based on the following known fact: if $\bm{X}$ is Gaussian, then the non-zeros in its precision matrix (or inverse covariance matrix) $\Omega_{0}:=\Sigma_{0}^{-1}$ corresponds exactly to links in the graphical model (see \cite{Hastie2015statistical,whittaker2009}). That is, there is a link between nodes $i$ and $j$ if and only if the $(i,j)$-th entry of $\Omega_{0}$ is non-zero. Similarly the lack of a link $\{i,j\}$ between a pair of nodes $i$ and $j$ is equivalent to the $(i,j)$-th entry of $\Omega_{0}$ being zero.

Now suppose we observe $T$ i.i.d samples ${\bm X}_{t}, t=1,\dots,T$ from $\bm{X} = (X_{1},\dots,X_{p})$. We now introduce the GLASSO estimator, as well as our proposed SGLASSO estimator, that allows us to recover the graphical model from the samples. The output of these estimators is a sparse precision matrix $\hat{\Omega}$. From  $\hat{\Omega}$, the estimated graphical model is then constructed by including a link between nodes $i$ and $j$ if and only if $\hat{\Omega}_{i,j} \neq 0$.




For a $m \times n$ matrix $A$, define the mixed $L_{p,q}$ norm as follows:
 
 \begin{align*}
 \Vert A \Vert_{p,q} 
=  \left[\sum_{j=1}^n \left( \sum_{i=1}^m |a_{ij}|^p \right)^{q/p}\right]^{1/q}.
   \end{align*}
   
   The following $L_{1,2}$ norm in equation (\ref{def:l12}) will play a crucial role in our estimator:
   
    \begin{align}
 \Vert A \Vert_{1,2} 
=  \left[\sum_{j=1}^n \left( \sum_{i=1}^m |a_{ij}| \right)^{2}\right]^{1/2}. \label{def:l12}
   \end{align}
   
   Our {\bf Structured-GLASSO} (SGLASSO) estimator is defined in equation (\ref{l12}) below, where $\hat{\Sigma}$ is the sample covariance matrix calculated from  the sample $({\bf X}_{t})_{t=1}^{T}$.
   
     \begin{align}
\hat{\Omega}_{\text{sglasso}} :=  \argmax_{\Omega\succ0} \Bigg\{&\log{\det \Omega} - \tr(\Omega \hat{\Sigma}) - \lambda_T \|\Omega\|^{2}_{1,2} \Bigg\}. \label{l12}
  \end{align}
  
 In comparison, the {\bf GLASSO} estimator (\cite{banerjee2008model,friedman2008sparse,yuan2007model,fan2015overview}) is defined in equation (\ref{l11}). While SGLASSO uses the $L_{1,2}$ mixed norm as a penalty to the likelihood, GLASSO in equation (\ref{l11}) uses the $L_{1,1}$ norm, which is the familiar lasso penalty (sum of the absolute values of the entries of $\Omega$). 
   
  \begin{align}
\hat{\Omega}_{\text{glasso}}  := \argmax_{\Omega\succ0} \Bigg\{&\log{\det \Omega} - \tr(\Omega \hat{\Sigma}) - \lambda_T \|\Omega\|_{1,1} \Bigg\}. \label{l11}
  \end{align}
    
Another definition of the GLASSO is one where only $\Omega^{-}$ is penalized, where $\Omega^{-}$ denotes the matrix $\Omega$ where its' diagonal entries are set to zero. While \cite{yuan2007model,rothman2008sparse} use equation (\ref{l11}), other authors such as \cite{banerjee2008model,friedman2008sparse} use the latter.  For various expositional reasons, we will penalize $\Omega$ instead of  $\Omega^{-}$. However in the Monte Carlo simulation, we will also compare with the variant of GLASSO where the diagonals are not penalized.

Although we are comparing SGLASSO with GLASSO in this paper, we could also combine them together. That is, $\hat{\Omega} := \argmax_{\Omega\succ0} \big\{\log{\det \Omega} - \tr(\Omega \hat{\Sigma}) - \lambda_{1T}\|\Omega\|_{1,1} - \lambda_{2T}\|\Omega\|^{2}_{1,2} \big\}$, and use cross-validation procedures to tune the parameters $\lambda_{1T}$ and $\lambda_{2T}$.

We do not require that the data-generating process be Gaussian.\footnote{When the DGP is non-Gaussian, the graphical model estimated using GLASSO or SGLASSO is still useful -- it corresponds to the structure of pairwise conditional correlations (partialling out other variables).} Therefore our estimator can be seen as maximizing a Gaussian quasi-likelihood subject to regularization.
When there is no penalty ($\lambda_T = 0$), then $\hat{\Omega}$ defined in both (\ref{l12}) and (\ref{l11}) correspond to the Quasi Maximum Likelihood estimator of the inverse covariance matrix, which is given by the inverse of the empirical sample covariance matrix. As we mentioned in the introduction, it is well-known that the unpenalized sample estimator behaves poorly, and it is unsuitable for graphical modeling when zeros of the estimates are important.

\subsection{Motivating Example}

We illustrate the difference between the $L_{1,2}$ and the conventional $L_{1,1}$ lasso norms. Define $\Omega_{1}$ and $\Omega_{2}$ as the precision matrices given by equations (\ref{o1}) and (\ref{o2}) below.

\begin{figure}[!htb]
    \centering
    \begin{minipage}{.5\textwidth}
        \centering
\begin{align}
\Omega_{1} = \begin{bmatrix}
       1 & \frac{1}{\sqrt{5}} & \frac{1}{\sqrt{5}}    & \frac{1}{\sqrt{5}} & \frac{1}{\sqrt{5}}      \\[0.2em]
      \frac{1}{\sqrt{5}} & 1 & 0    & 0 & 0       \\[0.2em]
     \frac{1}{\sqrt{5}} & 0 & 1    & 0 & 0       \\[0.2em]
      \frac{1}{\sqrt{5}} & 0 & 0    & 1 & 0       \\[0.2em]
       \frac{1}{\sqrt{5}} & 0 & 0    & 0 & 1       \\[0.2em]
     \end{bmatrix}\label{o1}
 \end{align}
    \end{minipage}%
    \begin{minipage}{0.5\textwidth}
        \centering
 \begin{align}
\Omega_{2} = \begin{bmatrix}
       1 & \frac{1}{\sqrt{5}} & 0    & 0 & 0      \\[0.2em]
      \frac{1}{\sqrt{5}} & 1 & \frac{1}{\sqrt{5}}    & 0 & 0       \\[0.2em]
     0 & \frac{1}{\sqrt{5}} & 1    & \frac{1}{\sqrt{5}} & 0       \\[0.2em]
      0 & 0 & \frac{1}{\sqrt{5}}   & 1 & \frac{1}{\sqrt{5}}       \\[0.2em]
       0 & 0 & 0    & \frac{1}{\sqrt{5}} & 1       \\[0.2em]
     \end{bmatrix}\label{o2}
 \end{align}
    \end{minipage}
\end{figure}

The precision matrices $\Omega_{1}$ and $\Omega_{2}$ give rise respectively to the graphical models in Figure \ref{fig:o1} and Figure \ref{fig:o2} below, where links in the graphs represent non-zero entries in the precision matrices. 
     
     \begin{figure}[!htb]
    \centering
    \begin{minipage}{.5\textwidth}
        \centering
        \includegraphics[scale=0.45]{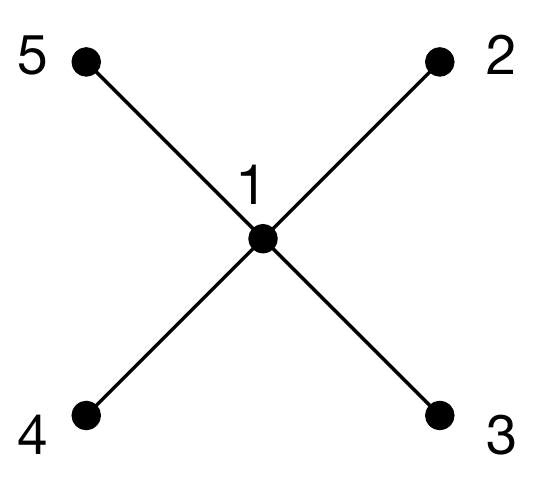}
        \caption{$\Omega_{1}$: star graph}
        \label{fig:o1}
    \end{minipage}%
    \begin{minipage}{0.5\textwidth}
        \centering
        \includegraphics[scale=0.4]{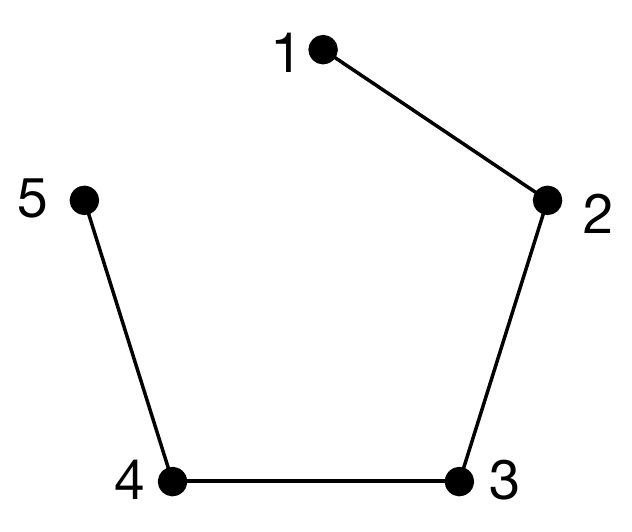}
        \caption{$\Omega_{2}$: AR(1) model.}
        \label{fig:o2}
    \end{minipage}
\end{figure}

Both the standard Lasso and Frobenius norms do not distinguish between $\Omega_{1}$ and $\Omega_{2}$, while our proposed  $L_{1,2}$-norm does. The reason is clear: both $\Omega_{1}$ and $\Omega_{2}$ have the same number of zeros. They have the same sparsity but the structure of this sparsity is very different. For the $L_{1,1}$ lasso norm, we have: $L_{1,1}(\Omega_{1}) = L_{1,1}(\Omega_{2}) = \sum_{j=1}^{p} \sum_{i \neq j}^{p}|\frac{1}{\sqrt{5}}| + 5 = \frac{8}{\sqrt{5}}+ 5$. Similarly, for the $L_{2,2}$ Frobenius norm, we have $L_{2,2}(\Omega_{1}) =  L_{2,2}(\Omega_{2})$.

 Now on the other hand, the $L_{1,2}$-norm can distinguish between these two structures. The $L_{1,2}$ norm evaluated at $\Omega_{1}$ for the star graph is $\|\Omega_{1}\|^{2}_{1,2}  \approx 16.16$.  While the $L_{1,2}$ norm evaluated at $\Omega_{2}$ for the AR(1) graph is $\|\Omega_{2}\|^{2}_{1,2} \approx 14.96$.  

\section{Theoretical properties}

In this section we derive the limiting distribution of SGLASSO when the number of samples $T$ goes to infinity while $p$ is fixed (following \cite{yuan2007model}). The main finding of this section is that SGLASSO is asymptotically equivalent to an infeasible GLASSO problem which prioritizes the sparsity-recovery of high-degree nodes.  More precisely, we show that
SGLASSO  is asymptotically equivalent to a modified GLASSO estimator where each row of $\Omega$ is penalized proportionally according to its {\em true} degree. The true degree of row (or node) $i$ is defined by $d_{i} = \sum_{j =1}^{p}|\Omega_{0,ij}|$. The term ``degree" should be understood as ``weighted degree" henceforth.

This asymptotic equivalence result has two implications. First, SGLASSO inherits the sparsity-discovery property of GLASSO where zeros in the precision matrix are estimated precisely to be zero (due to  the non-differentiability of the penalty function at zeros). Therefore SGLASSO estimates are also sparse. 
Secondly, SGLASSO differs from GLASSO in that it prioritizes recovering the zero-pattern of high-degree nodes. Nodes with higher weighted degrees can be viewed as being more influential and important, and we might be more interested in uncovering the relationships among influential agents.



To derive the limiting distributions of our SGLASSO estimator, we assume a high level assumption on the weak limit of the sample covariance matrix.

\begin{assumption}
	\label{as.limit.samplecovariance} We assume that 
	\[
	\text{vech}\: (\sqrt{T} (\hat{\Sigma} - \Sigma_0 )) \Rightarrow \text{vech}\: (W) \sim \mathcal{N}(0,\Lambda),
	\]
\end{assumption}
where the notation $vech$ is the half-vectorization operator that takes only the lower-triangular part of a symmetric matrix.

This high level condition holds when $X_{j,t} X_{k,t}$ have higher moments and the serial dependence is weak. When ${\mathbf X}_t \sim_{i.i.d.} \mathcal{N}(0,\Sigma_0)$, then  $\sqrt{T} ( \text{vech}\: (\hat{\Sigma} - \Sigma_0 )) \Rightarrow \text{vech} \: (W) \sim \mathcal{N}(0,\Lambda)$, where $\Lambda$ is such that $\Cov(W_{ij}, W_{i'j'}) = \Cov(X_{i,t}X_{j,t}, X_{i',t}X_{j',t})$.

Now let $D_p$ be a $p \times p$ matrix whose $(i,j)$ entry is $d_{j} \equiv \sum_{k=1}^{p} |\Omega_{0,jk}|$, the true weighted degree of node $j$. 
Define
\begin{equation}
\hat{\Omega}_{D_p} = \argmax_{\Omega \succ0} \{\log{\det \Omega} - \tr( \Omega \hat{\Sigma}) -  2\lambda_T \| D_p \circ \Omega\|_{1,1} \} \label{LASSO.equivalence}
\end{equation}
to be a variant of the GLASSO estimator. Here, the operator $\circ$ is the element-wise multiplication. The SGLASSO estimator is defined as: $\hat{\Omega}_{\lambda} = \argmax_{\Omega\succ0} \{\log{\det \Omega} - \tr(\Omega \hat{\Sigma}) - \lambda_{T} \| \Omega \|^{2}_{1,2} \}$.

\begin{theorem}\label{aequi}
Suppose that as $T \rightarrow \infty$, $\sqrt{T}\lambda_{T} \rightarrow \lambda_0 > 0$.  Then, 
\[
\sqrt{T}(\hat{\Omega}_{\lambda} - \Omega_0) = \sqrt{T}(\hat{\Omega}_{D_p} - \Omega_0) + o_p(1).  
\]
In addition, suppose that Assumption \ref{as.limit.samplecovariance} holds. Then,
\begin{align*}
\sqrt{T}(\hat{\Omega}_{\lambda} - \Omega_0), \text{ and } \sqrt{T}(\hat{\Omega}_{D_p} - \Omega_0)  \Rightarrow 
\argmin_{ U \in \mathbb{R}^{p \times p}} V(U), 
\end{align*}
where 
\begin{align}
V(U) = \tr(U \Sigma_{0} U \Sigma_{0}) + \tr(U W) + 2 \lambda_{0}\sum_{i=1}^{p} \sum_{j=1}^{p} g_{ij}(u_{ij})d_{j} \label{VU}
\end{align}
as $T \rightarrow \infty$. Here $\Sigma_{0} = \Omega_{0}^{-1}$ is the true covariance matrix of $X$, and $g_{ij}(u):= u\, \text{sign}(\Omega_{0,ij})\mathds{1}(\Omega_{0,ij} \neq 0) + |u|\mathds{1}(\Omega_{0,ij} = 0)$. 
\end{theorem}

Compared to the limit in Theorem \ref{aequi}, the GLASSO estimator has the following limiting distribution (\cite{yuan2007model})\footnote{\cite{yuan2007model} consider the GLASSO estimator where the diagonals are not penalized, so that $V_{\text{glasso}}(U) = \tr(U \Sigma_{0} U \Sigma_{0}) + \tr(U W) + \lambda_{0} \sum_{i=1}^{p} \sum^{p}_{j \neq i} g(u_{ij})$ in their paper.}:  
	\[
	\argmin_{U \in \mathbb{R}^{p \times p}} V_{\text{glasso}}(U)
	\] 
	where
	\begin{align}
	V_{glasso}(U) = \tr(U \Sigma_{0} U \Sigma_{0}) + \tr(U W) + \lambda_{0} \sum_{i=1}^{p} \sum^{p}_{j = 1} g_{ij}(u_{ij}) \label{vglasso}
	\end{align}
	where $\sqrt{T}\lambda_T \rightarrow \lambda_0$. 

Comparing equations (\ref{VU}) and (\ref{vglasso}), we see that the SGLASSO obtains the same limiting distribution as that of a modified GLASSO with element-wise LASSO penalty given by $2\| D_p \circ \Omega\|_{1,1}$. Since $D_p$ is constructed using the true, unknown values of $\Omega_{0}$, SGLASSO cannot be simply implemented as GLASSO.

Moreover from Proposition \ref{prop:didj} below, we can further say that SGLASSO is asymptotically equivalent to a variant of GLASSO where the penalty term is $\sum_{i=1}^p \sum_{j=1}^p (d_i + d_j) | \Omega_{ij} |$. That is, each entry $\Omega_{ij}$ is given the penalty factor equals to the sum of the (true) weighted degrees of nodes $i$ and $j$.

\begin{proposition}\label{prop:didj}
Let $D_p$ be a $p \times p$ matrix whose $(i,j)$ entry is $d_{j} \equiv \sum_{k=1}^{p} |\Omega_{0,jk}|$. It is true that $2 \| D_p \circ \Omega\|_{1,1}  = \sum_{i=1}^p \sum_{j=1}^p (d_i + d_j) | \Omega_{ij} |$.  
\end{proposition}

All proofs are relegated to the Appendix. In the Supporting Information, we plot and illustrate the asymptotic distribution in Theorem \ref{aequi}. We show that with a higher probability, SGLASSO correctly estimates a non-link (a zero entry) belonging to a high-degree node.


The distribution of $\argmin_{ U \in \mathbb{R}^{p \times p}}V(U)$ can be simulated after replacing the unknown components $\Sigma_0, \Omega_0,$ and $\Lambda$ with their consistent estimates, and this can be used in inference procedures on $\Omega_0$. 

A noticeable feature, however, is that the limit distribution in Theorem \ref{aequi} is discontinuous with respect to true DGP. This is because $g_{ij}(u)$ is discontinuous as a function of $\Omega_{0,ij}$ at $\Omega_{0,ij}=0$. This implies that the inference based on this limit would only be valid pointwise, not uniformly in $\Omega_0$.\footnote{See \cite{belloni2014inference} and \cite{van2014asymptotically}. We thank one of the referees who pointed this out.}

\section{Simulation Results}

\begin{figure}[!h]
\centering
\begin{subfigure}{0.32\textwidth}
\centering
\includegraphics[scale=0.1]{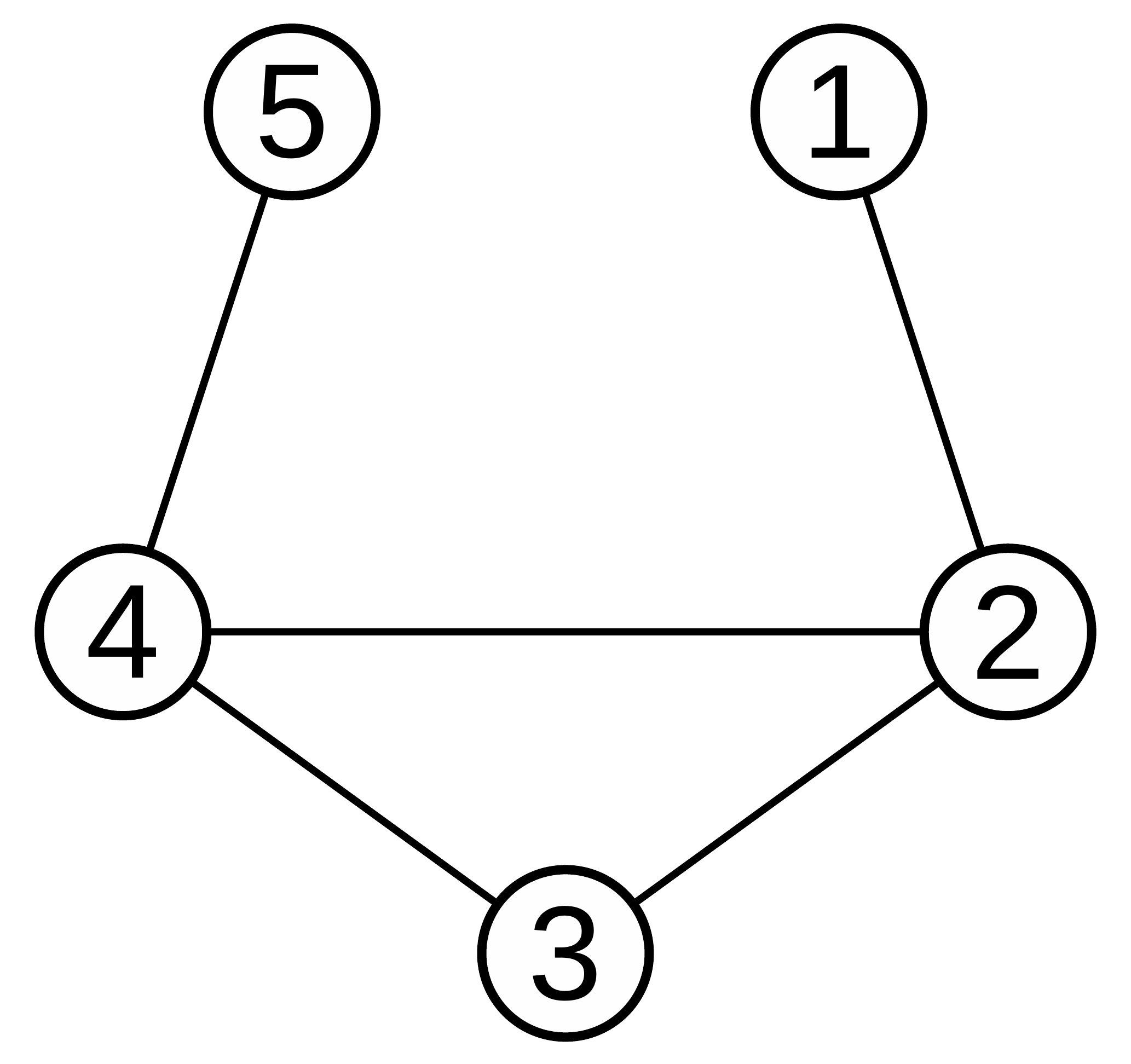}
\caption{Model 1}
\end{subfigure} 
\begin{subfigure}{0.32\textwidth}
\centering
\includegraphics[scale=0.15]{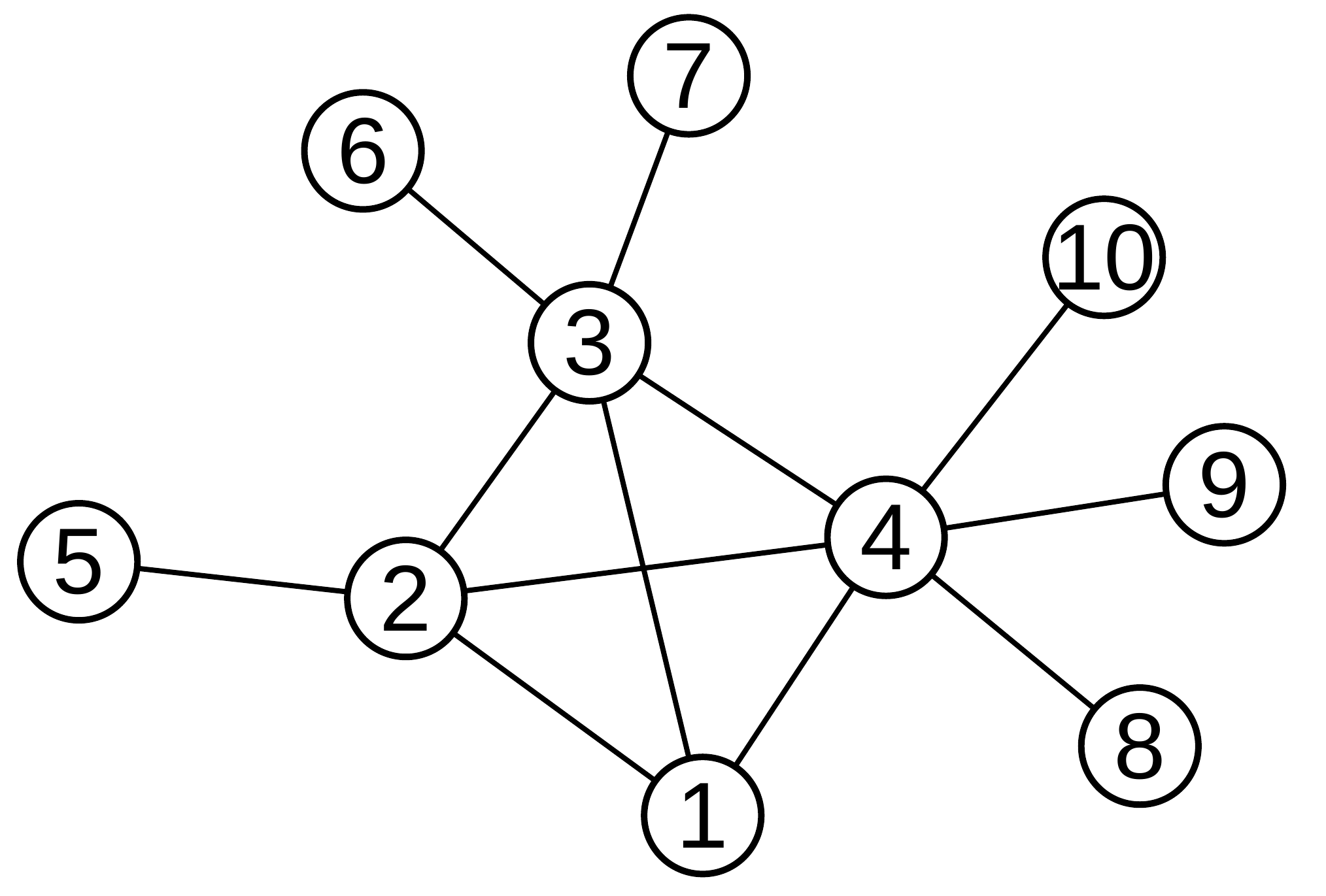}
\caption{Model 10}
\end{subfigure}
\caption{$p=5$ and $=10$}\label{fig:model110}
\end{figure}

We compare our estimator against GLASSO by considering 3 different graphical models here. In the Supplemental Information, we consider 8 other models. The first two models are depicted in Figure \ref{fig:model110}. The last model is the graphical model calibrated to the empirical application as depicted in Figure \ref{fig:firm_sglasso}. From a given graphical model, we generate the true precision matrix $\Omega_{0}$ such that $\Omega_{0,ij} = 0$ if and only if there is a link between nodes $i$ and $j$, otherwise we set $\Omega_{0,ij} = 0.2$. We set $\Omega_{0,ii} = 1$.

For each model, we draw 1,000 independent datasets from $\mathcal{N}(0,\Omega^{-1}_{0})$. That is, each dataset comprises of  $({\bf X}_{t})_{t=1}^{T}$, where ${\bf X}_{t} \in \mathbb{R}^{p}$ is randomly drawn from  $\mathcal{N}(0,\Omega^{-1}_{0})$. For the sample size, we consider $T=20$  and $T=50$.  


Both estimators involve choosing the $\lambda_T$ tuning parameters. We use a 2-folds cross-validation procedure to tune $\lambda_T$. Specifically, we use the Kullback-Leibler (KL) loss averaged over the two-folds to evaluate predictive accuracies. Equation (\ref{KLcv}) gives the KL loss between the estimated $\hat{\Omega}$ from the training set versus the estimated $\Omega$ from the validation set:
\begin{align}
KL(\lambda_{T}) = \log \det(\Omega) - \log \det \hat{\Omega} + \tr(\hat{\Omega} \Omega^{-1}) -  p \label{KLcv}
\end{align}

We report the simulation result in Table \ref{cvmc1}. In the table, the (a) columns corresponds to SGLASSO whereas the (b) columns refer to GLASSO. In Columns 1(a) and 1(b), we report the optimal $\lambda_T$ as determined by cross-validations, averaged across the 1,000 replications. In Columns 2(a) and 2(b), we report the Kullback-Leibler loss averaged across 1,000 replications.\footnote{The KL loss between $\hat{\Omega}$ and $\Omega_{0}$ is given by $KL(\hat{\Omega},\Omega_{0}) = \log \det(\Omega_{0}) - \log \det \hat{\Omega} + \tr(\hat{\Omega} \Omega_{0}^{-1}) -  p$} In Columns 3(a) and 3(b), we report the average Frobenius loss between $\hat{\Omega}$ and $\Omega_{0}$. 

 In the last two columns of Table \ref{cvmc1}, we report the $F_{1}$ score, which measures the accuracy of graph recovery. The Kullback-Leibler loss and the Frobenius norms may not fully capture how accurately the zeros are recovered. We introduce an additional metric:  $F_{1} = \frac{2 \text{precision}\cdot \text{recall}}{\text{precision}+\text{recall}}$, where {\em precision} is the ratio of true positives (TP) to all predicted positives (TP + FP), {\em recall} is the ratio of true positives to all actual positives (TP + FN).  Alternatively, the $F_{1}$ score can be written as $F_{1} = \frac{2TP}{2TP + FP + FN}$. The $F_{1}$ score measures the quality of a binary classifier by equally balancing both the precision and the recall of a classifier. The larger the $F_{1}$ score is, the better the classifier is. The $F_{1}$ score is commonly used in machine learning to evaluate binary classifiers. For instance, the Yelp competition uses the $F_1$ score as a metric to rank competing models.\footnote{https://www.kaggle.com/c/yelp-restaurant-photo-classification} The $F_{1}$ score is favored over the metric $Accuracy = \frac{TP+TN}{TP + TN + FP + FN}$ especially in our current setting where the graphical models are sparse. This is because a model that naively predicts all negatives will obtain a high {\em Accuracy} score just because there are many actual negatives, and the TN term dominates the {\em Accuracy} score.

\begin{table}[!h]
{\small
\makebox[\textwidth][c]{
\begin{tabu}{ccccccccccccc}
\toprule
Model & $T$ & \multicolumn{2}{c}{Optimal $\lambda_{T}$} && \multicolumn{2}{c}{KL}  & & \multicolumn{2}{c}{Frobenius} & & \multicolumn{2}{c}{$F_{1}$ score}  \\
\cmidrule{3-4} \cmidrule{6-7} \cmidrule{9-10} \cmidrule{12-13}
& & (a)  & (b)  && (a) & (b) && (a) & (b)  && (a) & (b)  \\
& & {\scriptsize SGLASSO}  & {\scriptsize GLASSO}  && {\scriptsize SGLASSO} &  {\scriptsize GLASSO} && {\scriptsize SGLASSO} &  {\scriptsize GLASSO}  && {\scriptsize SGLASSO} &  {\scriptsize GLASSO} \\
\hline
(1) &$20$ & 0.181 & 0.342 && 0.430 & 0.483  && 0.896 & 0.948 && 0.439 & 0.367  \\
\rowfont{\footnotesize} 
& & (0.118) & (0.133) && (0.207) & (0.215) && (0.223) & (0.214) && (0.248) & (0.259)  \\ 
(10) &$20$& 0.207 & 0.399 && 0.946 & 1.096  && 1.353 & 1.456 && 0.336 & 0.272  \\ 
\rowfont{\footnotesize}
 && (0.094) & (0.109) && (0.278) & (0.527) && (0.183) & (0.414) && (0.143) & (0.162)  \\ 
  (11) & 20 & 0.198 & 0.389 && 1.201 & 1.390  && 1.569 & 1.701 && 0.403  & 0.341  \\ 
\rowfont{\footnotesize}
 & & (0.097) & (0.117) && (0.312) & (0.723) && (0.186) & (0.522) && (0.143) & (0.164)  \\ 
 \hline
 (1) &$50$ & 0.088 & 0.179 && 0.246 & 0.265  && 0.710 & 0.736 && 0.564 & 0.541  \\ 
\rowfont{\footnotesize} 
 && (0.054) & (0.076) && (0.084) & (0.091) && (0.129) & (0.134) && (0.225) & (0.248)  \\ 
 (10) &$50$ & 0.109 & 0.230 && 0.593 & 0.651  && 1.137 & 1.190 && 0.466 & 0.437  \\ 
\rowfont{\footnotesize}
 && (0.040) & (0.052) && (0.115) & (0.128) && (0.113) & (0.117) && (0.136) & (0.159)  \\ 
  (11)& 50  & 0.091 & 0.206  && 0.747 & 0.807  && 1.312 & 1.370 && 0.563 & 0.541  \\ 
\rowfont{\footnotesize}
& & (0.041) & (0.053) && (0.153) & (0.167) && (0.128) & (0.133) && (0.120) & (0.137)  \\ 
\bottomrule
\end{tabu}
}
\caption{Averages and standard errors from 1,000 replications.}\label{cvmc1}}
\end{table}

 The conclusion from Table \ref{cvmc1} is that SGLASSO achieves significantly lower Kullback-Leibler and Frobenius losses. Moreover in in terms of graph accuracy, SGLASSO also outperforms GLASSO, as indicated by the $F_{1}$ scores.

 To abstract away from the effects of cross-validations, we consider the lowest Kullback-Leibler losses that can be achieved by our estimator versus the GLASSO. Specifically, we vary $\lambda_{T}$ and at each $\lambda_{T}$, we compute the KL and Frobenius losses between the true $\Omega_{0}$ and $\hat{\Omega}$. We see from Table \ref{kltable} that the superior performance of SGLASSO over GLASSO exists after taking away the randomness due to cross-validations. The lowest possible KL losses achievable by our estimator appears to be smaller, than the corresponding KL losses for the GLASSO.

\begin{table}[h!]
{\small
\centering
\begin{tabu}{ccccccccc}
\toprule
Model & $T$& \multicolumn{2}{c}{Minimum}  & & \multicolumn{2}{c}{Minimum} & & \multicolumn{1}{c}{Percentage}  \\
& & \multicolumn{2}{c}{KL}  & & \multicolumn{2}{c}{Frobenius} & & \multicolumn{1}{c}{Dominance}  \\
\cmidrule{3-4} \cmidrule{6-7} 
& & (a)  & (b)  && (a) & (b) &&  \\
& & {\footnotesize SGLASSO}  & {\footnotesize GLASSO}  && {\footnotesize SGLASSO} &  {\footnotesize GLASSO} &&  \\
 \hline
(1) & 20 & 0.340 & 0.376 && 0.779 & 0.821  && 0.99 \\ 
& & (0.106) & (0.120) && (0.096) & (0.116)  & \\ 
(10) & 20 & 0.824 & 0.887 && 1.234 & 1.279  && 0.99 \\ 
& & (0.162) & (0.176) && (0.085) & (0.101)  & \\ 
(11) &20 & 1.041 & 1.102 && 1.415 & 1.466  && 0.95 \\ 
& & (0.191) & (0.205) && (0.087) & (0.101)  & \\ 
 \hline
 (1) & 50 & 0.199 & 0.211 && 0.606 & 0.634  && 0.92 \\ 
 & & (0.063) & (0.065) && (0.095) & (0.098)  & \\ 
 (10)& 50 & 0.509 & 0.533 && 1.021 & 1.035  && 0.96 \\ 
& & (0.087) & (0.095) && (0.072) & (0.082)  & \\ 
(11) & 50 & 0.652 & 0.675 && 1.203 & 1.204  && 0.86 \\ 
 & & (0.104) & (0.114) && (0.067) & (0.082)  & \\ 
\bottomrule
\end{tabu}
\caption{Minimum possible KL and Frobenius losses across all $\lambda_T$ (averages and standard errors from 1,000 replications). In the last column, we report the fraction of times in which SGLASSO performed better in terms of KL loss at the corresponding $\lambda_T$.}
\label{kltable}
}
\end{table}

Theorem \ref{aequi} says that the SGLASSO prioritizes recovering the sparsity between nodes that have higher degrees. We now show some numerical evidence that lends support to this. For each model, we examined the pair of nodes $(i,j)$ such that $d_{i} + d_{j}$ is largest and $\Omega_{0,ij} = 0$, where $d_{i} = \sum_{k=1}^{p}|\Omega_{0,ik}|$ is the true weighted degree of node $i$. For instance, for model 1, the $\argmax_{i,j} (d_{i}+d_{j})$ such that $\Omega_{0,ij} =0$ would correspond to the pairs of nodes $(1,4)$ and $(2,5)$. For simplicity, if there are multiple pairs of nodes that maximize $d_{i}+d_{j}$, we will pick the pair of nodes that comes first when the adjacency matrix is vectorized. 

In each of the model (set $T=20$), we calculate the fraction of times that GLASSO and SGLASSO correctly recover $\Omega_{0,ij} =0$ for the largest $d_{i} + d_{j}$. We plot this result in Figure \ref{fig:prob}, which shows that at any given $\lambda$, the probability of correctly recovering $\Omega_{0,ij} =0$ for high-degree nodes is greater when SGLASSO is used, compared to GLASSO.

\begin{figure}[!h]
\centering
\begin{subfigure}{0.3\textwidth}
\centering
\includegraphics[scale=0.3]{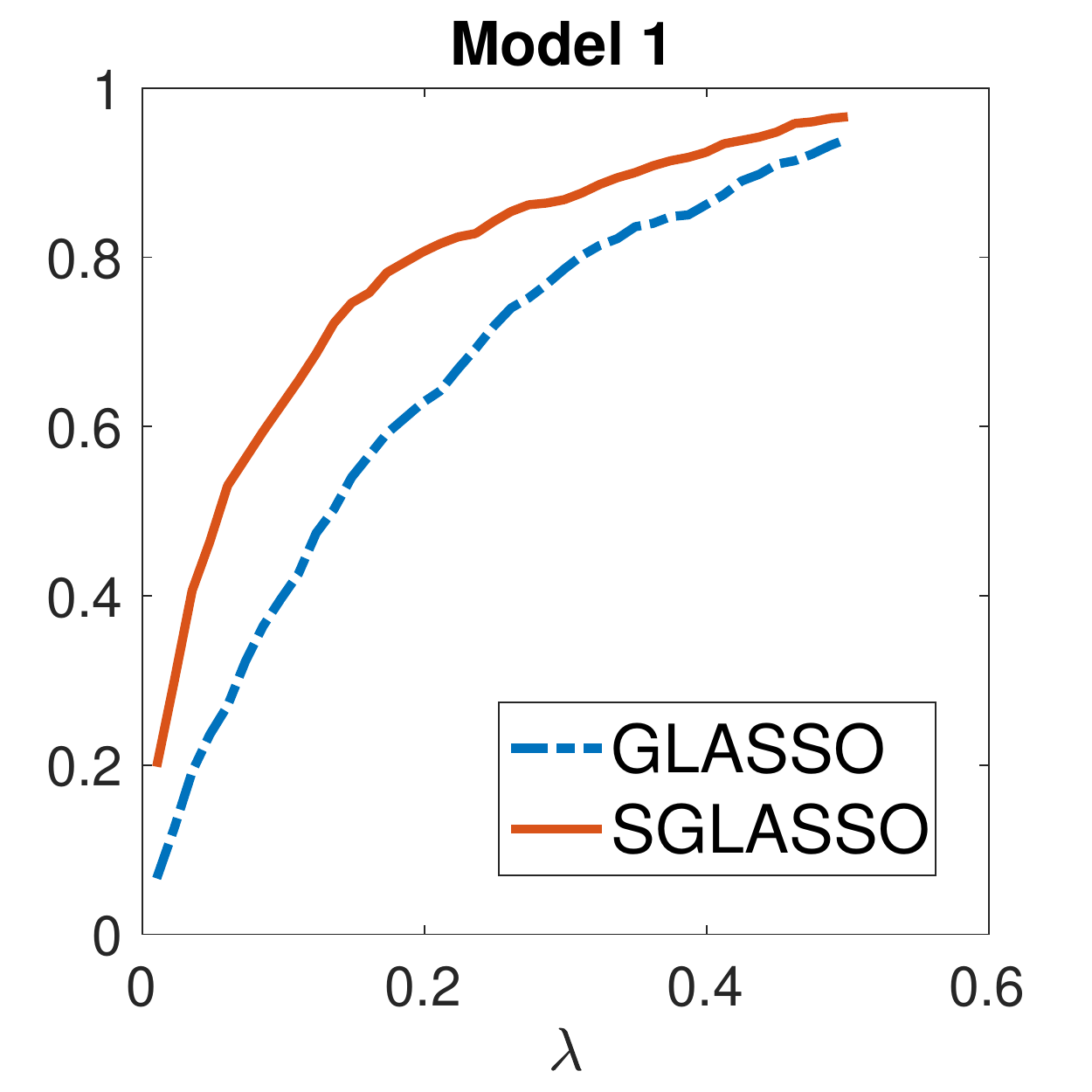}
\caption{Model 1}
\end{subfigure} 
\begin{subfigure}{0.3\textwidth}
\includegraphics[scale=0.3]{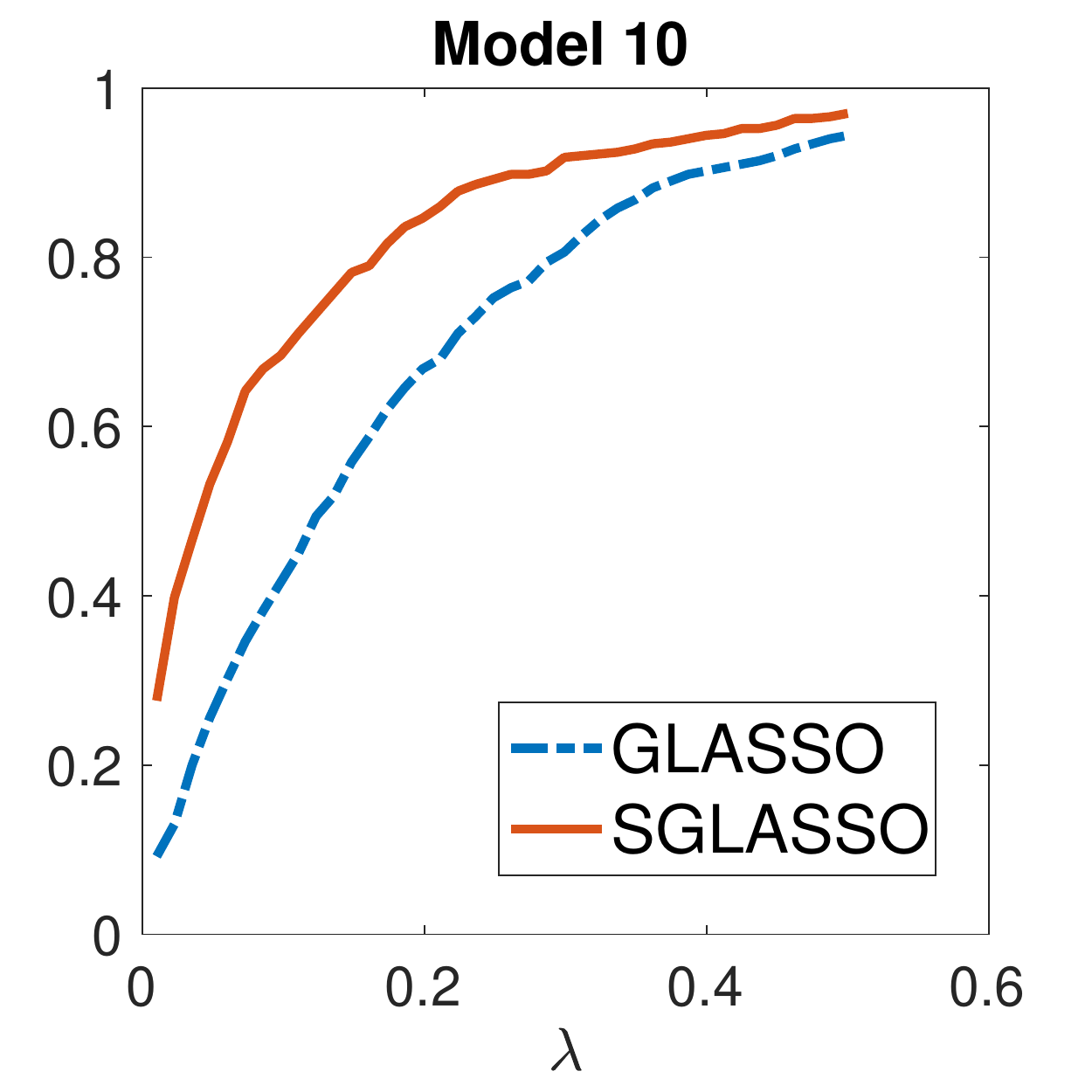}
\caption{Model 10}
\end{subfigure}
\begin{subfigure}{0.3\textwidth}
\includegraphics[scale=0.3]{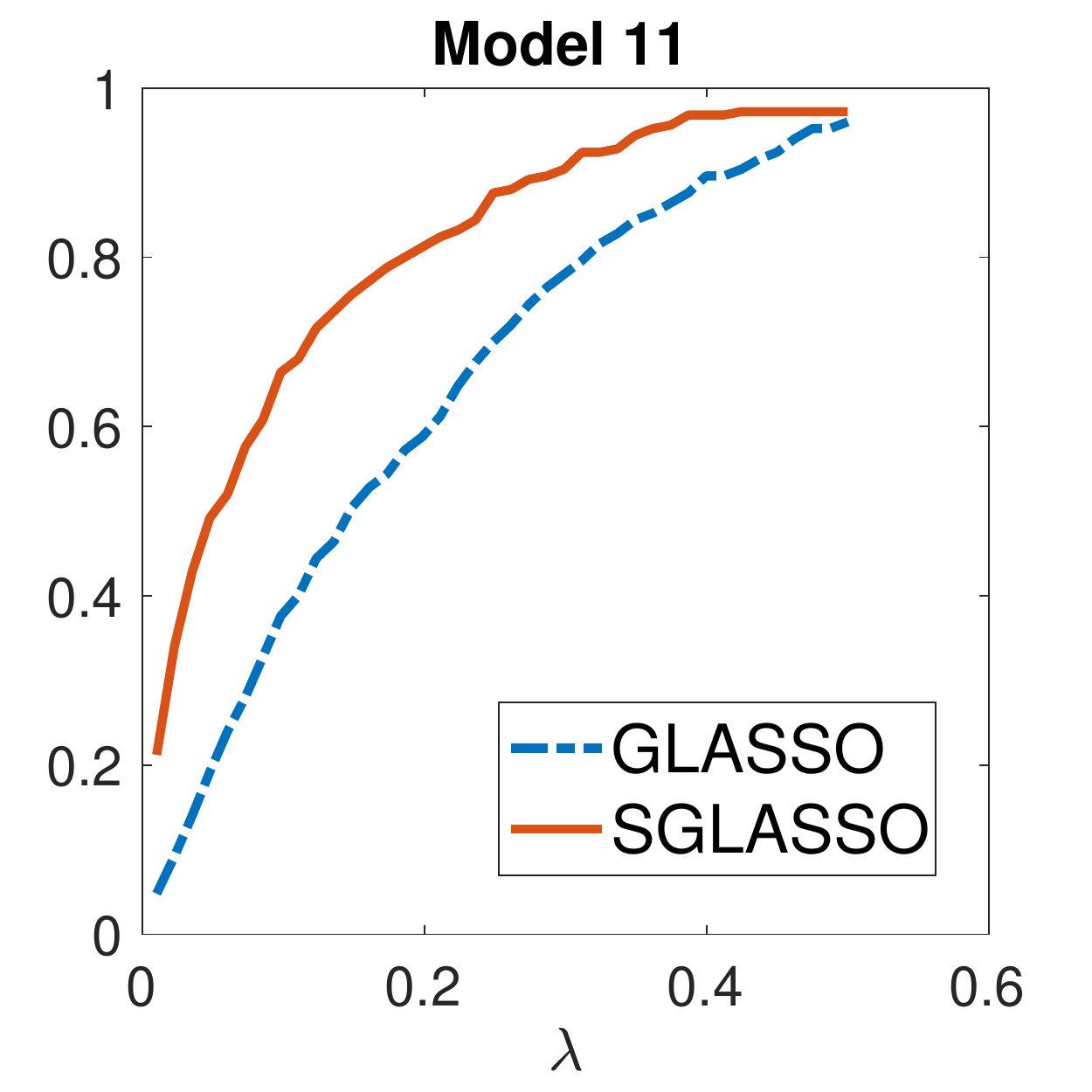}
\caption{Model 11}
\end{subfigure}
\caption{Each figure shows the estimated probability of recovering the non-link $(i,j)$ such that $(d_{i} + d_{j})$ is the largest in each model, as a function of the tuning parameter $\lambda_T$.}\label{fig:prob}
\end{figure}

\subsection{Computational details}

Computing the SGLASSO estimator is a convex optimization problem. More precisely, the problem consists of minimizing a smooth convex term given by $-\log{\det \Omega} + \tr(\Omega \hat{\Sigma})$, and a non-differentiable convex term corresponding to $\lambda_T\|\Omega\|^{2}_{1,2}$. Since the objective can be formulated as minimizing the sum of a differentiable convex function and a non-differentiable convex function, we can use the {\em proximal gradient method} (\cite{beck2009fast}) to compute SGLASSO.

In this paper, we use CVX, a Matlab package for specifying and solving convex programs (\cite{grant2008cvx}). CVX supports convex objective functions that are non-smooth (\cite{grant2008graph}). CVX also allows us to easily enforce symmetry and positive-definiteness of the matrix minimand, which is needed here. The code is readily available from the authors upon request.

\section{Empirical illustration}

For a real-world empirical illustration, we use the classic Grunfeld investment data (see \cite{greene2012econometric} or \cite{baltagi2008econometric}). The goal is to estimate the network dependence structure of firms' investment decision.\footnote{The data can be downloaded freely from \cite{greene2012econometric}'s companion website.}


The data consist of time series of $T = 20$ yearly observations on $p=10$ firms.  The three variables are $I_{it} = $ real gross investment of firm $i$ in year $t$, $F_{it} = \text{real market value of the firm}$, $C_{it} = \text{real value of capital stock such as plant and equipment}$. 

The investment decision for firm $i$ is modeled as follows: $I_{it} = \beta_{1i} + \beta_{2i}F_{it} + \beta_{3i}C_{it} + \epsilon_{it}$, where $(\epsilon_{1t},\epsilon_{2t},\dots,\epsilon_{pt})  \sim iid  (0,\Omega_{0}^{-1})$. The zeros and non-zeros in $\Omega_{0}$ correspond to the dependence structure of firms' investment decisions (the object of interest here).

\begin{figure}[!h]
\centering
\includegraphics[scale=0.65]{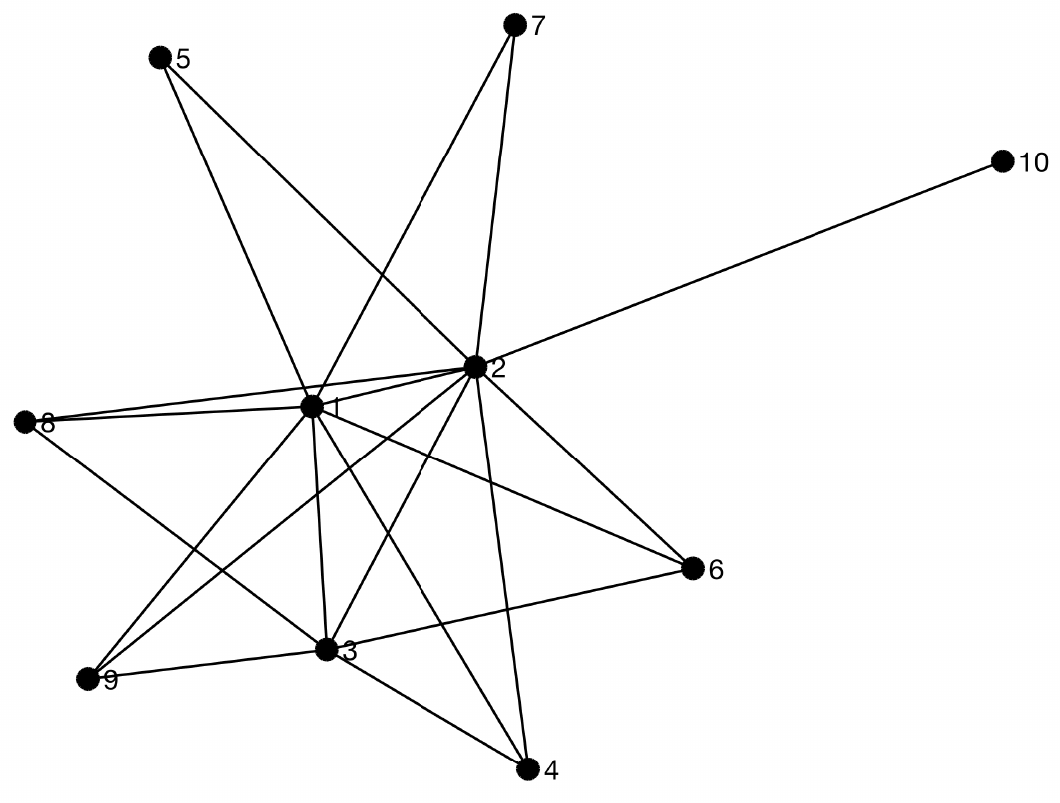}
\caption{Network of firms' dependence estimated using SGLASSO. The estimated graphical model exhibits a clear core-periphery structure. The core firms are 1, 2 and 3, which are respectively General Motors, U.S. Steel and General Electric. The periphery firms do not have links to each other, but are linked only to the core firms. All core firms are linked to each other.}\label{fig:firm_sglasso}
\end{figure}

For each firm $i$, we first estimate the linear equation: $I_{it} = \beta_{1i} + \beta_{2i}F_{it} + \beta_{3i}C_{it} + \epsilon_{it}$. This is also the first step in the estimation of Seemingly Unrelated Regressions. Secondly, we use our proposed SGLASSO  to estimate the precision matrix of the first-step residuals $\hat{\epsilon}_{it}$, $i=1,\dots,10$, $t=1,\dots,20$. The tuning parameter is determined using a two-fold cross-validation procedure. The result is shown in Figure \ref{fig:firm_sglasso}. (We also estimated the graphical model corresponding to the observed investment variable $I_{it}$, the result is similar)

The recovered graphical model exhibits a clear core-periphery structure. There are two groups of firms: a group of core firms and a group of periphery firms. The core firms are linked to each other, as well as linked to the periphery firms. The periphery firms however, are {\em not} linked to each other, but are  only linked to the core firms. Therefore every node is linked directly to a core firm. The core firms here are firms 1, 2 and 3, which are respectively General Motors, U.S. Steel and General Electric. Using the GLASSO estimator, we recover an identical graphical model -- the finding is robust to different choices of penalty functions. 

Finally we cannot recover this core-periphery structure using the sample estimator (without  regularization). To show this, we estimate the sample precision matrix of the residual estimates $\hat{\epsilon}_{it}$ from the first step. The graphical model corresponding to this sample precision matrix is the complete graph, where every node is linked to all other nodes -- the sample precision matrix is dense and do not contain any zeros.

\section{Concluding remarks and policy implications}

Graphical model is a potentially useful tool for economists. Using graphical models, we can obtain the network of dependence among random variables. We introduce Structured Graphical Lasso (SGLASSO) as an estimator of graphical models. Using a classic firms' investment dataset, we find that the dependence structure of firms' investment decision exhibits a core-periphery network. This has some relevant policy implications -- small shocks to those core firms will affect the entire network, creating large aggregate fluctuations as in \cite{acemoglu2012network}. Indeed per our finding, one of the core firms is General Motors, whose Chapter 11 restructuring posed a systemic risk to the U.S. economy in 2009.

%



\section*{Acknowledgements}

We thank Victor Chernozhukov and two referees for helpful comments and suggestions. The first draft of the paper was written while Moon and Chiong were Associate Director and a postdoctoral fellow of USC Dornsife INET, respectively. Moon acknowledges that this work was supported by the Ministry of Education of the Republic of Korea and the National Research Foundation of Korea (NRF-2017S1A5A2A01023679).

\appendix

\section*{Appendix}
\section{Proofs}

{\small

We will use the following important result that concerns the asymptotics of the minimizer of a convex random function (\cite{geyer1994asymptotics,hjort2011asymptotics,kato2009asymptotics,pollard1991asymptotics}).

\begin{lemma}\label{convexlemma}
Let $f_{n}({\bf x},\Omega): \mathbb{R}^{d} \times \Omega \rightarrow \mathbb{R}$ be a random convex function, that is, $f_{n}({\bf x},\cdot)$ is a random variable for each ${\bf x} \in \mathbb{R}^{d}$. Moreover, let ${\bf x}^{*}_{n}(\Omega)$ be the unique minimizer of $f_{n}({\bf x},\Omega)$ with respect to ${\bf x}$. Suppose that $f_{n}({\bf x},\cdot)$ converges  in distribution to some random convex function $f_{\infty}({\bf x},\cdot)$ for each ${\bf x}$. Now if ${\bf x}^{*}_{\infty}(\Omega)$ is the unique minimizer of $f_{\infty}({\bf x},\Omega)$, then the random variable ${\bf x}^{*}_{n}$ converges in distribution to the random variable ${\bf x}^{*}_{\infty}$
\end{lemma}

\subsection{Proof of Theorem \ref{aequi}}

To prove Theorem \ref{aequi}, define the following convex random functions by re-parameterization $U = \sqrt{T} (\Omega - \Omega_0)$: 

\begin{align}
V_{T}(U) =& -\log \det\left(\Omega_0 + \frac{U}{\sqrt{T}}\right) + \tr \left\{ \left(\Omega_0 + \frac{U}{\sqrt{T}}\right)\hat{\Sigma}\right \}  
+ \lambda_{T} \bigg\|\Omega_{0} + \frac{U}{\sqrt{T}}\bigg\|^{2}_{1,2} \notag \\
& + \log \det \Omega_0 - \tr(\Omega_0 \hat{\Sigma}) -  \lambda_{T} \|\Omega_{0}  \|^{2}_{1,2}, \label{VTU}
\end{align}
and 
\begin{align}
V_{T}^l(U) =& -\log \det\left(\Omega_0 + \frac{U}{\sqrt{T}}\right) + \tr \left\{ \left(\Omega_0 + \frac{U}{\sqrt{T}}\right)\hat{\Sigma}\right \}  
+ 2\lambda_{T} \bigg\|D_p \circ \left(\Omega_0 + \frac{U}{\sqrt{T}} \right)\bigg\|_{1,1} \notag \\
& + \log \det \Omega_0 - \tr(\Omega_0 \hat{\Sigma}) -  2 \lambda_{T} \|D_p \circ \Omega_{0} \|_{1,1}. \label{VTU.lasso}
\end{align}

This function is random because of its dependence on the sample covariance matrix, $\hat{\Sigma}$. By definition, our SGLASSO estimator  $\hat{\Omega}_{\lambda}$ satisfies the following.
\[
\sqrt{T} (\hat{\Omega}_{\lambda} - \Omega_0) = \hat{U} := \argmin_{U} V_{T}(U).
\]

By Lemma \ref{convexlemma}, the limit distribution of $\sqrt{T} (\hat{\Omega}_{\lambda} - \Omega_0)$ follows if we show 
\begin{equation}
T V_{T}(U) \rightarrow V(U), \label{eq.theorem2.required}
\end{equation}
as $T \rightarrow \infty$ and as $\sqrt{T} \lambda_{T} \rightarrow \lambda_{0} \geq 0$, where the random limit function $V(U)$ is defined in equation (\ref{VU}).

Following the argument in the proof of Theorem 1 of \cite{yuan2007model}, we have that:
\begin{eqnarray}
\log \det\left(\Omega_0 + \frac{U}{\sqrt{T}}\right) - \log \det \Omega_0 
&=& \frac{\tr(U\Sigma_0)}{\sqrt{T}}-  \frac{\tr(U\Sigma_0 U\Sigma_0)}{T} + o\left(\frac{1}{T}\right) \label{Tlogdet} \\
\tr \left\{ \left(\Omega_0 + \frac{U}{\sqrt{T}}\right)\hat{\Sigma}\right \} - \tr(\Omega_0 \hat{\Sigma}) 
&=& \tr \left(\frac{U \hat{\Sigma}}{\sqrt{T}} \right) \nonumber \\
&=& \tr \left(\frac{U\Sigma_0}{\sqrt{T}}\right) + \tr\left( \frac{U(\hat{\Sigma} - \Sigma_0)}{\sqrt{T}}\right) \label{Ttrace}
\end{eqnarray}
Moreover, we know that when $T$ is large enough, 
\begin{align*}
\left| \Omega_{0,ij} + \frac{u_{ij}}{\sqrt{T}}\right | - \left| \Omega_{0,ij} \right | = \frac{g_{ij}}{\sqrt{T}}
\end{align*}
where $g_{ij} = u_{ij}\, \text{sign}(\Omega_{0,ij})\mathds{1}(\Omega_{0,ij} \neq 0) + |u_{ij}|\mathds{1}(\Omega_{0,ij} = 0)$. 
Therefore we can show that the difference of the penalty terms can be written as
\begin{align}
\left\| \Omega_{0} + \frac{U}{\sqrt{T}}\right \|^{2}_{1,2} - \left\| \Omega_{0} \right \|^{2}_{1,2}  
&=\sum_{j=1}^{p} \left(\sum_{i =1}^{p} \left|\Omega_{0,ij} + \frac{u_{ij}}{\sqrt{T}}\right|\right)^{2} - \sum_{j=1}^{p} \left(\sum_{i =1}^{p} \left|\Omega_{0,ij} \right|\right)^{2} \notag \\
&= \sum_{j=1}^{p}\left[\sum_{i =1}^{p} \left(\left|\Omega_{0,ij} + \frac{u_{ij}}{\sqrt{T}}\right| + \left|\Omega_{0,ij} \right|\right)\sum_{i =1}^{p} \left(\left|\Omega_{0,ij} + \frac{u_{ij}}{\sqrt{T}}\right| - \left|\Omega_{0,ij} \right|\right)\right]\notag \\
&=\sum_{j=1}^{p}\left[\sum_{i =1}^{p} \left(\left|\Omega_{0,ij} + \frac{u_{ij}}{\sqrt{T}}\right| + \left|\Omega_{0,ij} \right|\right)\sum_{i =1}^{p}\frac{1}{\sqrt{T}} g_{ij}\right] \notag \\
&=\sum_{j=1}^{p}\left[\sum_{i =1}^{p} \left(\left|\Omega_{0,ij} + \frac{u_{ij}}{\sqrt{T}}\right| - \left|\Omega_{0,ij} \right| +2\left |\Omega_{0,ij} \right | \right)\sum_{i =1}^{p}\frac{1}{\sqrt{T}} g_{ij}\right] \notag  \\
&=\sum_{j=1}^{p}\left[\left(\sum_{i =1}^{p} \frac{1}{\sqrt{T}}g_{ij}
 +\sum_{i =1}^{p}2\left |\Omega_{0,ij} \right | \right)\sum_{i =1}^{p}\frac{1}{\sqrt{T}} g_{ij}\right]  \notag \\
 &=\frac{1}{T}\sum_{j=1}^{p}\sum_{i,i' =1}^{p}g_{ij}g_{i'j}
 +\frac{2}{\sqrt{T}}\sum_{j=1}^{p}\sum_{i,i'  =1}^{p}\left |\Omega_{0,ij} \right | g_{i'j} \label{gij}
\end{align}
  
We can rewrite the second term of equation (\ref{gij}) as follows:
\begin{align*}
\sum_{j=1}^{p}\sum_{i,i'  =1}^{p} g_{ij}\left |\Omega_{0,i'j} \right |  
&= \sum_{j=1}^{p}\sum_{i,i'  =1}^{p} g_{ij}\left |\Omega_{0,ji'} \right | \\
&= \sum_{j=1}^{p}\sum_{i =1}^{p} g_{ij}\left(\sum_{i' =1}^{p} \left |\Omega_{0,ji'} \right |\right) \\
&=\sum_{i = 1}^{p} \sum_{j =1}^{p} g_{ij}d_{j},
\end{align*}
where $d_{j}$ is the true weighted degree of node $j$ defined as $d_{j} \equiv \sum_{k=1}^{p} |\Omega_{0,jk}|$. 

Combining equations (\ref{VTU}), (\ref{Tlogdet}), (\ref{Ttrace}) and (\ref{gij}), we can write $T V_{T}(U)$ as 
\begin{align*}
TV_{T}(U)  &= \tr(U\Sigma_{0} U\Sigma_{0})  + \tr\left(U \sqrt{T}(\hat{\Sigma} - \Sigma_0)\right) \\
&\quad \quad +\lambda_{T} \sum_{j=1}^{p}\sum_{i,i' =1}^{p}g_{ij}g_{i'j}
 +2\sqrt{T} \lambda_{T} \sum_{i=1}^{p} \sum_{j=1}^{p} g_{ij}d_{j} +o(1) 
 \end{align*}
In the limit as $T \rightarrow \infty$, we then have:
 \begin{align*}
TV_{T}(U)  = \tr(U\Sigma_{0} U\Sigma_{0})  + \tr\left(U \sqrt{T}(\hat{\Sigma} - \Sigma_0)\right) 
+2\lambda_{0}\sum_{i=1}^{p} \sum_{j=1}^{p} g_{ij}d_{j} 
\end{align*}
Similarly, in the limit as $T \rightarrow \infty$, we have:
\[
T V_{T}^l(U) = \tr(U\Sigma_{0} U\Sigma_{0})  + \tr\left(U \sqrt{T}(\hat{\Sigma} - \Sigma_0)\right) 
+2\lambda_{0}\sum_{i=1}^{p} \sum_{j=1}^{p} g_{ij}d_{j}
\]

Therefore, we have the first result of the theorem.

Under Assumption \ref{as.limit.samplecovariance} the second result of the theorem follows by the continuous mapping theorem and Lemma \ref{convexlemma}. $\blacksquare$

\subsection{Proof of Proposition \ref{prop:didj}}

\begin{proof}
Since $\Omega$ is symmetric, we have 
	\begin{align}
	2 \| D_p \circ \Omega\|_{1,1} &= 2 \sum_{i=1}^p \sum_{j=1}^p d_j |\Omega_{ij}| \\
	&= 2 \sum_{i=2}^p \sum_{j<i}^p d_j |\Omega_{ij}| +  2 \sum_{j=1}^{p-1} \sum_{i>j}^pd_i |\Omega_{ji}| + 2 \sum_{i=1}^p  d_i |\Omega_{ii}| \nonumber \\
	&= 2 \sum_{i=2}^p \sum_{j<i}^p d_j |\Omega_{ij}| +  2 \sum_{j=1}^{p-1} \sum_{i>j}^p d_i |\Omega_{ij}| + 2 \sum_{i=1}^p  d_i |\Omega_{ii}| \nonumber \\
	&= 2 \sum_{i=2}^p \sum_{j<i}^p d_j |\Omega_{ij}| +  2 \sum_{i=2}^p \sum_{j<i}^p d_i |\Omega_{ij}| + 2 \sum_{i=1}^p  d_i |\Omega_{ii}| \nonumber \\
		&= 2 \sum_{i=2}^p \sum_{j<i}^p (d_{i} + d_j) |\Omega_{ij}| +  \sum_{i=1}^p  (d_i+d_{i}) |\Omega_{ii}| \nonumber \\
	& = \sum_{i=1}^p \sum_{j=1}^p (d_i + d_j) | \Omega_{ij} |
	\label{LASSO.equivalence.penality}
	\end{align}
	\end{proof}

\section{Illustrating the asymptotic distribution}

We consider the following true precision matrix, $\Omega_{0}$. It is an AR(1) model with $p=4$. The corresponding graph representing $\Omega_{0}$ is depicted in Figure \ref{square}.

\begin{figure}[!htb]
	\centering
	\begin{minipage}{.5\textwidth}
		\centering
		\begin{align*}
		\Omega_0  = 
		\begin{pmatrix}
		1 & 0.5 & 0 & 0 \\
		0.5 & 1 & 0.5 & 0 \\
		0  & 0.5  & 1 & 0.5  \\
		0 & 0 & 0.5 & 1  
		\end{pmatrix}
		\end{align*}
	\end{minipage}
	\begin{minipage}{0.48\textwidth}
		\centering
		\includegraphics[scale=0.5]{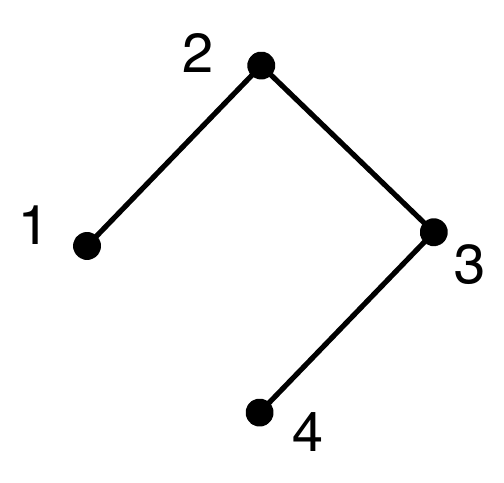}
		\caption{Graph of $\Omega_0$} \label{square}
	\end{minipage}
\end{figure}

In Figure \ref{l12plot} here, we illustrate the asymptotic distribution for both  GLASSO and SGLASSO, setting $\lambda_{0}=1$ in Theorem \ref{aequi} (see the main text). Each circle in the plot represents a draw from the distribution over $\hat{\Omega}$ as given by Theorem \ref{aequi}. Note that the precision matrix is such that $\Omega_{0,13} = \Omega_{0,14} =0$.  For GLASSO, we have $\hat{\pr}(\hat{\Omega}_{13} = 0) = 0.039$, while for SGLASSO, $\hat{\pr}(\hat{\Omega}_{13} = 0) = 0.124$. Therefore, SGLASSO does a better job at estimating $\Omega_{0,13} = 0$, where a non-link involves a node of relatively higher degree.

\begin{figure}[!htb]
	\centering
	\begin{minipage}{.5\textwidth}
		\centering
		\includegraphics[scale=0.4]{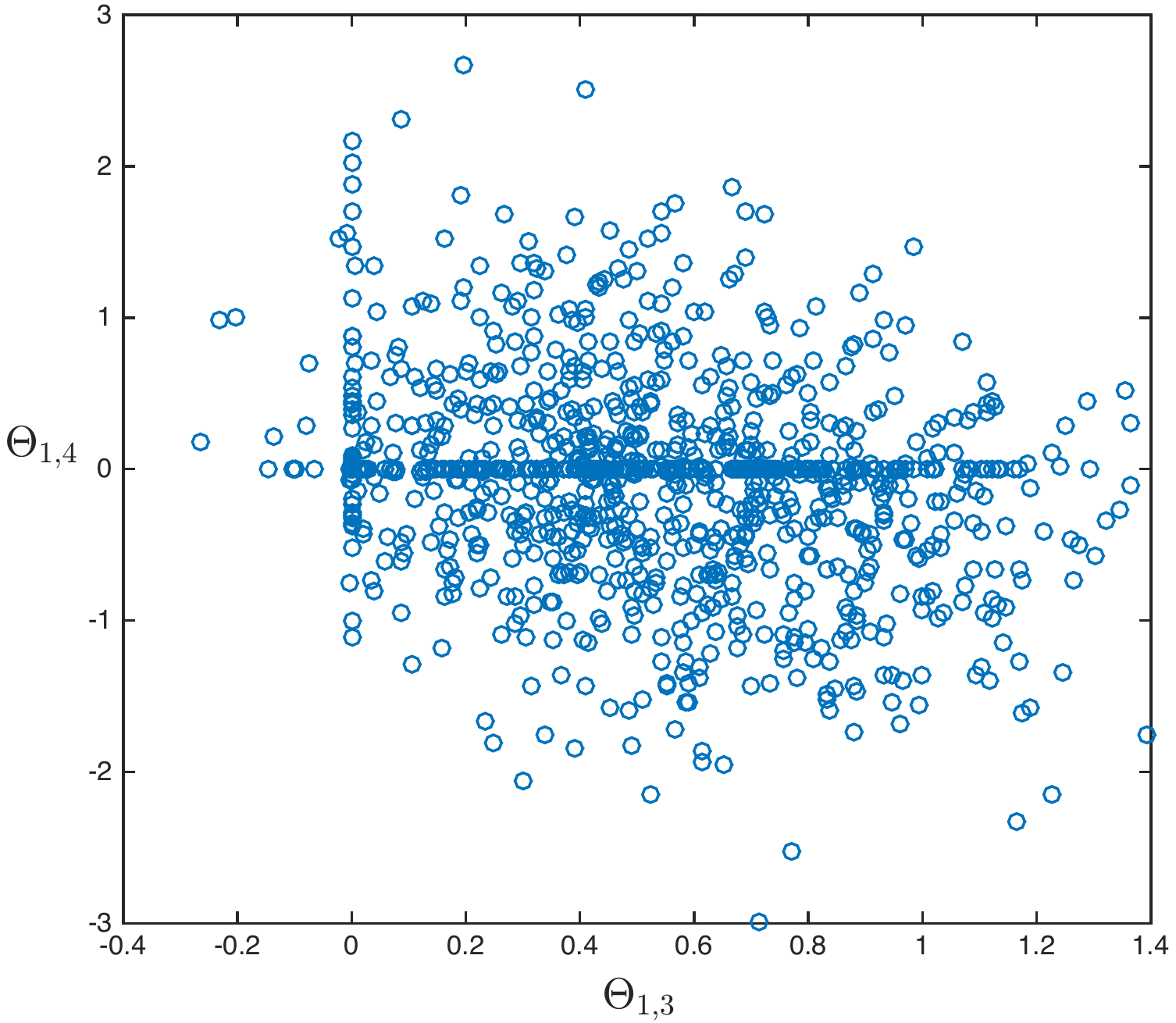}
	\end{minipage}%
	\begin{minipage}{0.5\textwidth}
		\centering
		\includegraphics[scale=0.4]{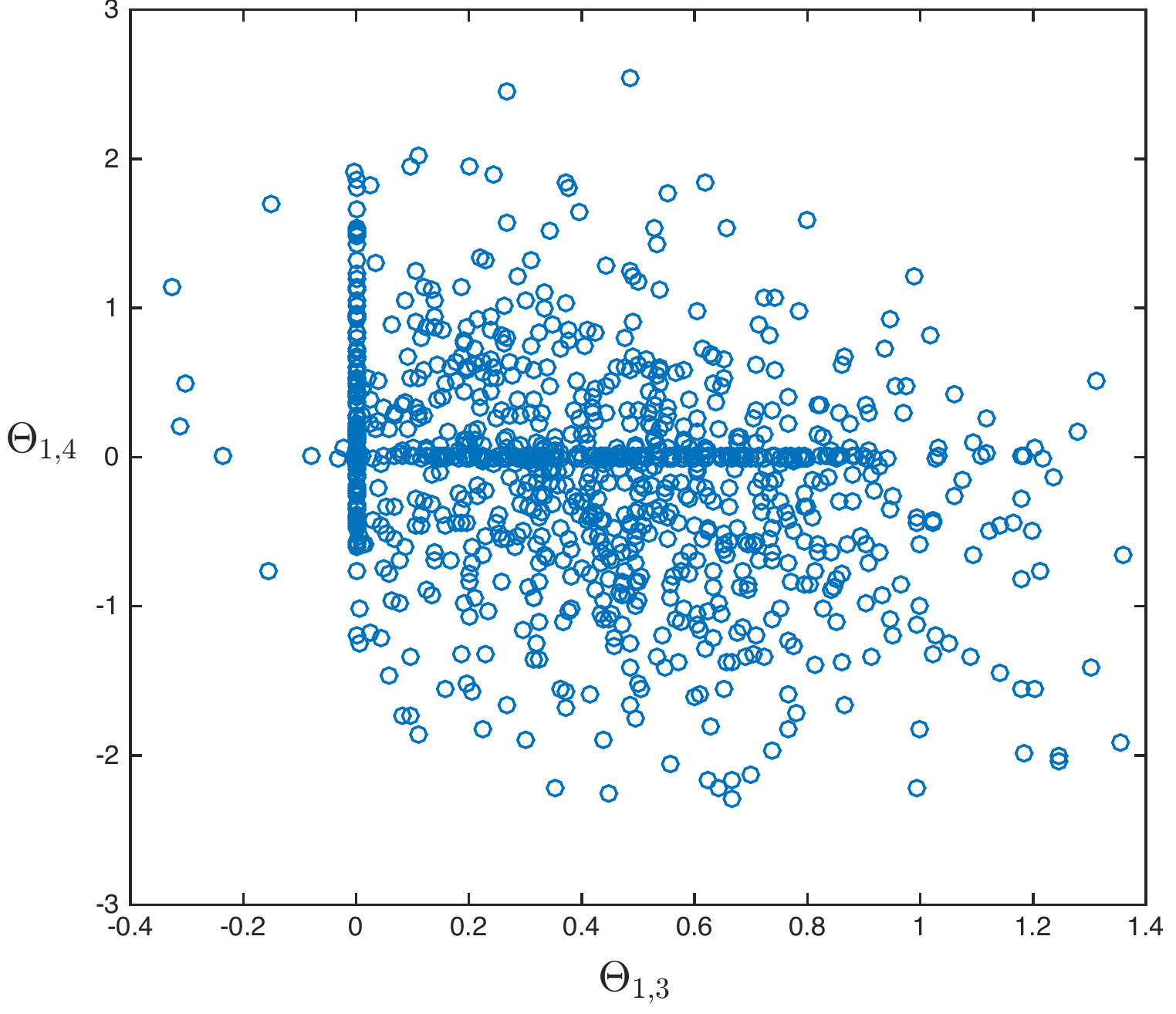}
	\end{minipage}
	\caption{Asymptotic distributions, with $\lambda_{0}=1$. Left-hand side is GLASSO, right-hand side is SGLASSO. For GLASSO,  $\hat{\pr}(\hat{\Omega}_{13} = 0) = 0.039$, $\hat{\pr}(\hat{\Omega}_{14} = 0) = 0.19$. For SGLASSO, $\hat{\pr}(\hat{\Omega}_{13} = 0) = 0.124$, $\hat{\pr}(\hat{\Omega}_{14} = 0) = 0.173$. Standard deviations of both estimators are comparable.} \label{l12plot}
\end{figure}

In Figure \ref{mle}, we  show the asymptotic distribution of $\hat{\Omega}$ without any penalty or regularization, which corresponds to the Maximum-Likelihood estimator of the inverse covariance matrix. Comparing the two figures, we see that both $L_{1,1}$ and $L_{1,2}$ norms are able to estimate entries of $\Omega_0$ to be zero exactly, while MLE without regularization fails to recover any sparsity structure.

\begin{figure}[!htb]
	\centering
	\includegraphics[scale=0.3]{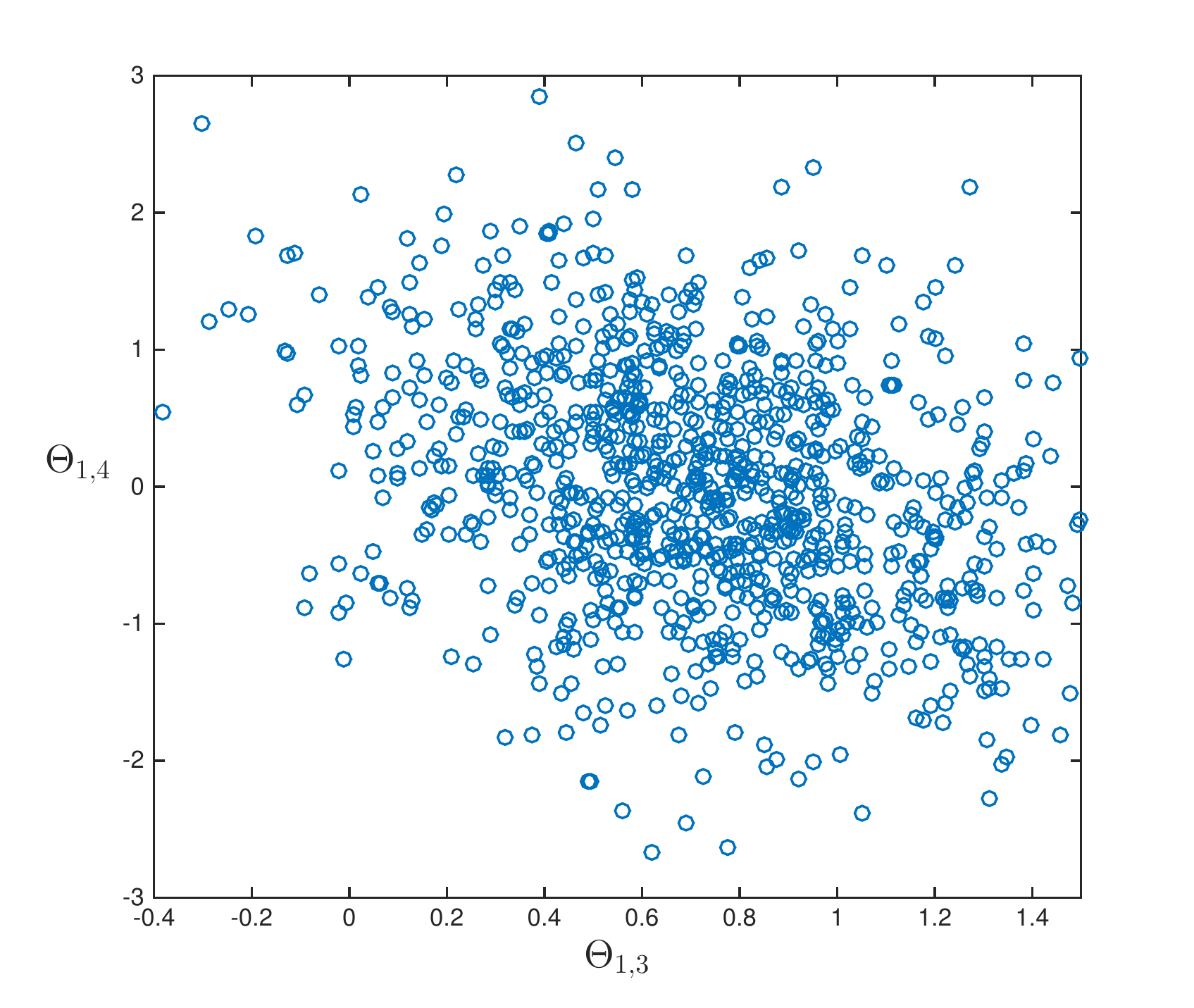}
	\caption{Asymptotic distribution of the Maximum Likelihood Estimator (MLE). This corresponds to the case where $\lambda_{0} = 0$.} \label{mle}
\end{figure}

%

\section{Simulation results}

In addition to the 3 graphical models considered in the main text, we explore 8 other models here.  The first 10 different graphical models are as depicted in Figures \ref{fig:p5} and \ref{fig:p10}. The last model is the graphical model calibrated to the empirical application. The models in Figure \ref{fig:p5} has 5 nodes ($p=5$), and the models in Figure \ref{fig:p10} has 10 nodes ($p=10$). From a given graphical model, we generate the true precision matrix $\Omega_{0}$ such that $\Omega_{0,ij} = 0$ if and only if there is a link between nodes $i$ and $j$, otherwise we set $\Omega_{0,ij} = 0.2$. We set $\Omega_{0,ii} = 1$.

For each model, we draw 1,000 independent datasets from $N(0,\Omega^{-1}_{0})$. That is, each dataset comprises of  $({\bf X}_{t})_{t=1}^{T}$, where ${\bf X}_{t} \in \mathbb{R}^{p}$ is randomly drawn from  $N(0,\Omega^{-1}_{0})$. Here, our sample size is $T=20$ when $p=5$ and $T=50$ when $p=10$.  For each of the 100 dataset, we use our estimator, as well as the GLASSO, to estimate the underlying inverse covariance matrix.

\begin{figure}[!h]
	\centering
	\begin{subfigure}{0.32\textwidth}
		\centering
		\includegraphics[scale=0.1]{model1.pdf}
		\caption{Model 1}
	\end{subfigure} 
	\begin{subfigure}{0.32\textwidth}
		\centering
		\includegraphics[scale=0.1]{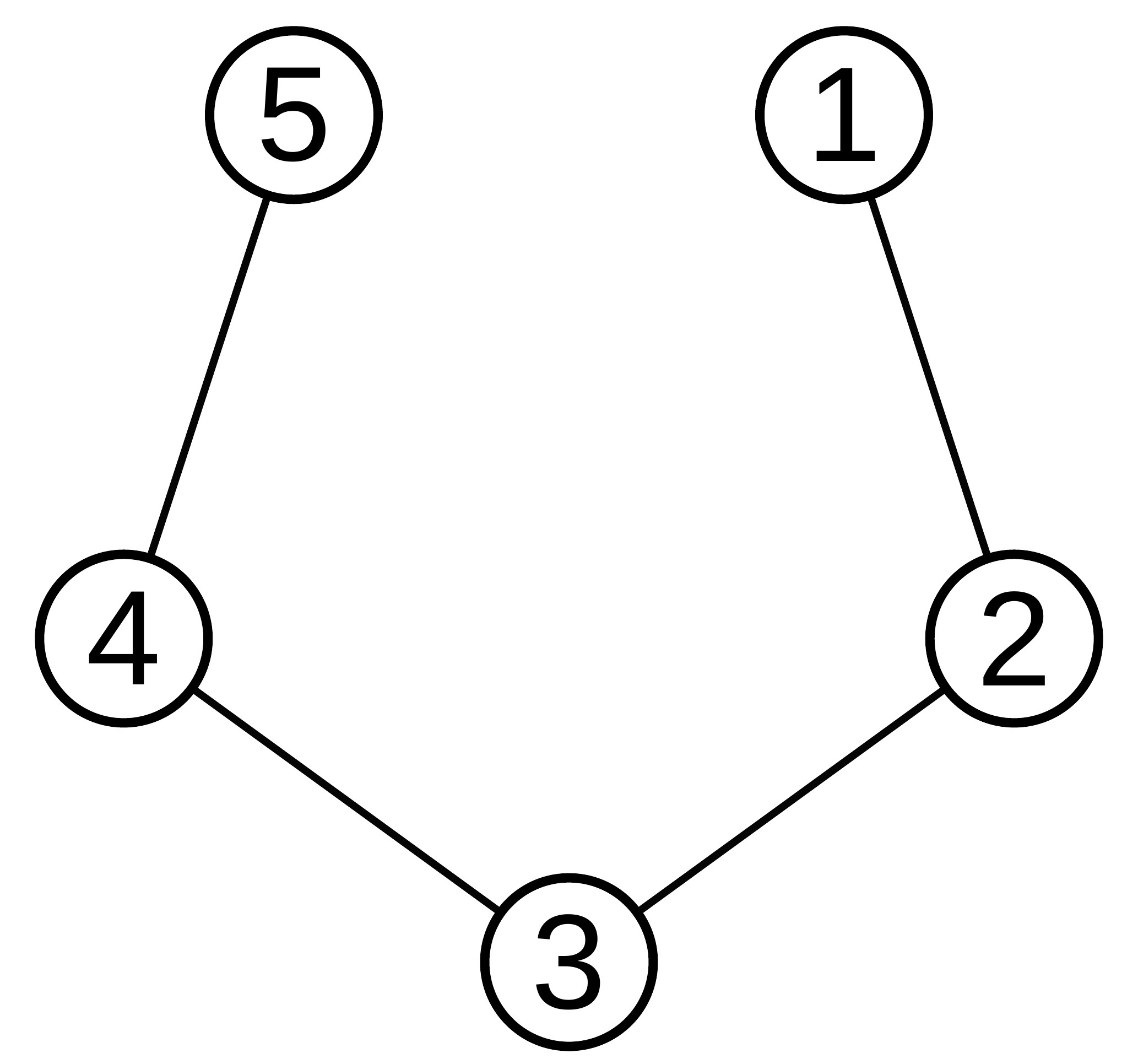}
		\caption{Model 2}
	\end{subfigure}
	\begin{subfigure}{0.32\textwidth}
		\centering
		\includegraphics[scale=0.1]{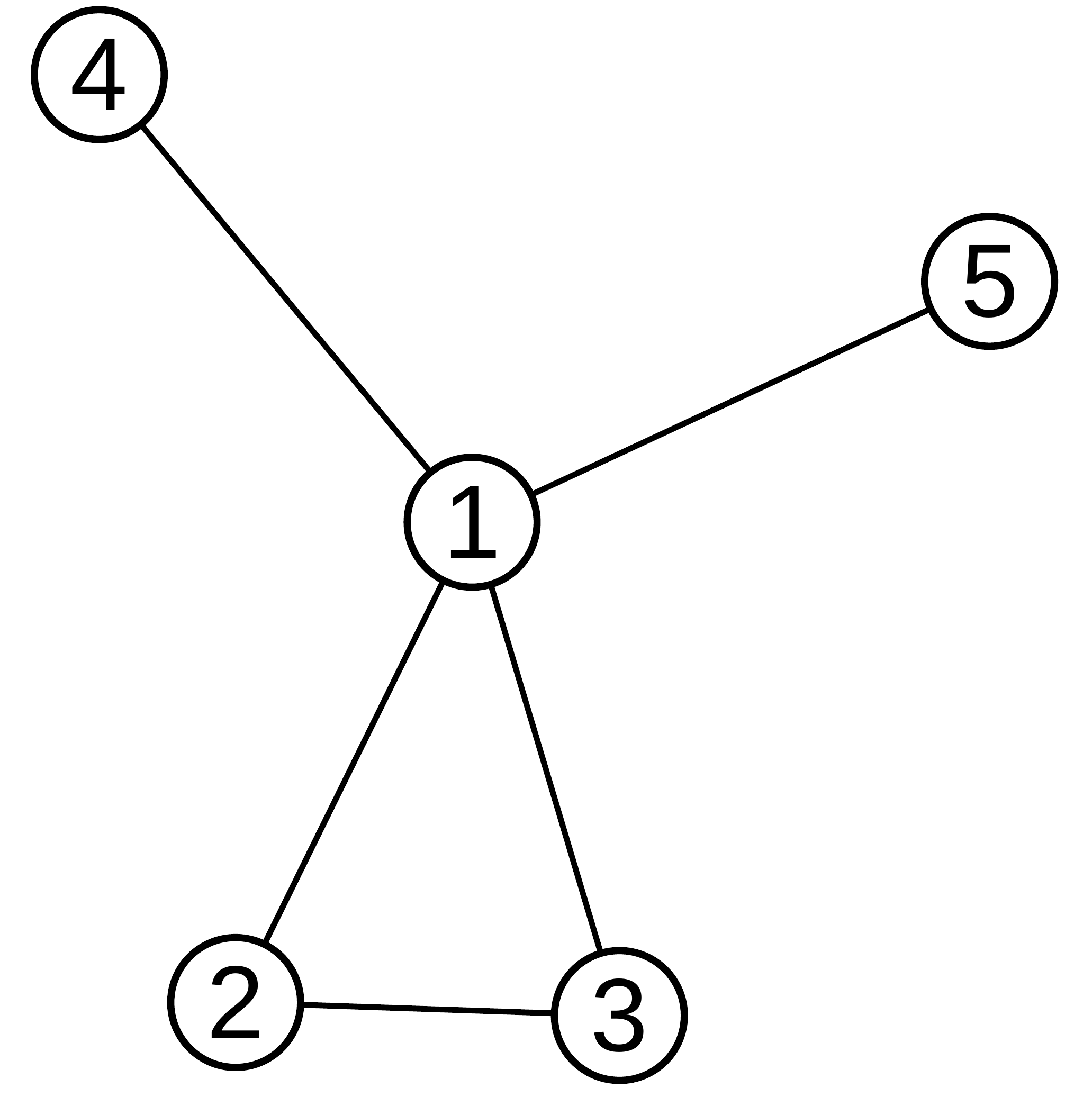}
		\caption{Model 3}
	\end{subfigure}
	
	\hspace{0.15\textwidth}
	\begin{subfigure}{0.32\textwidth}
		\centering
		\includegraphics[scale=0.1]{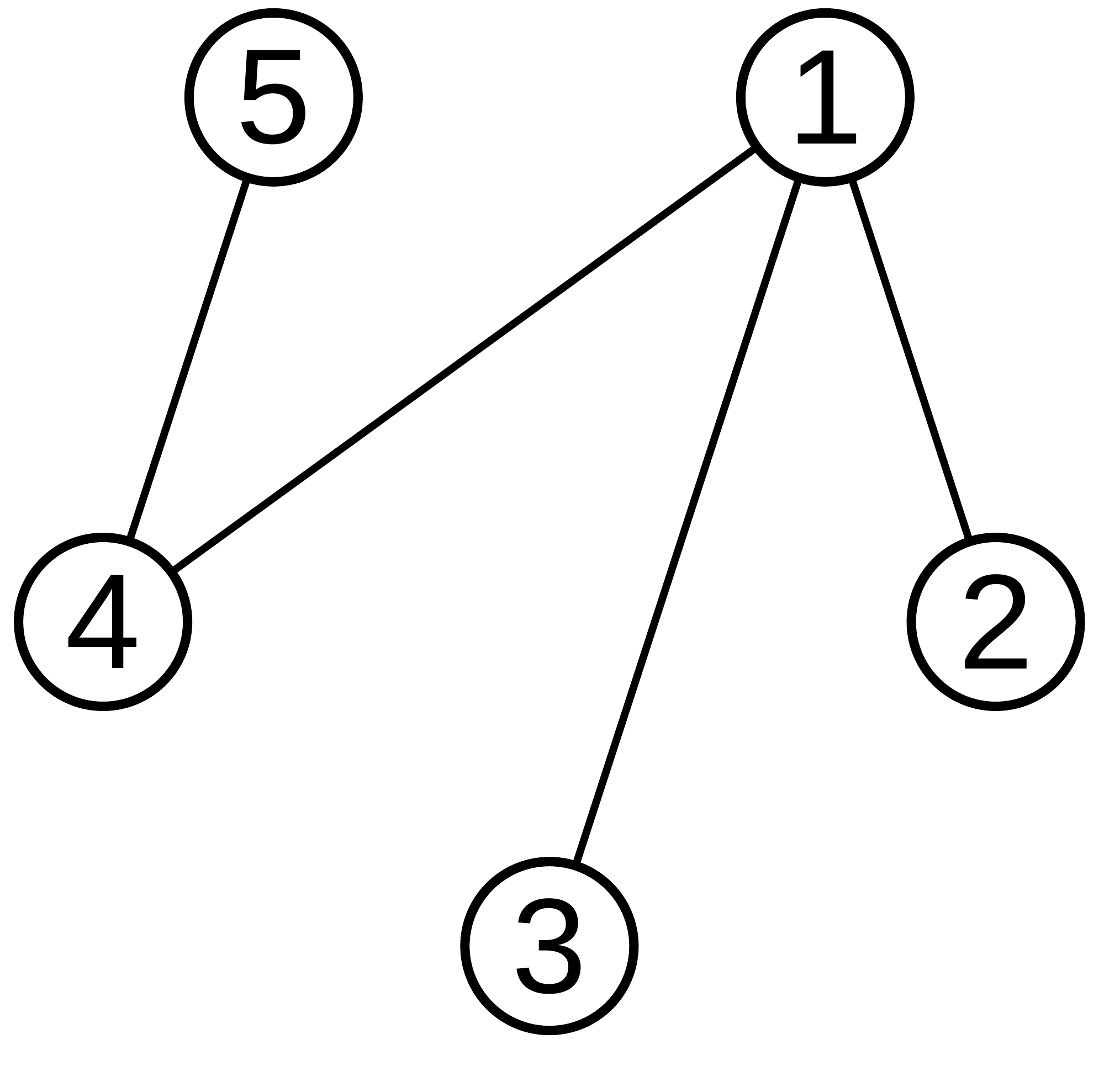}
		\caption{Model 4}
	\end{subfigure}
	\begin{subfigure}{0.32\textwidth}
		\centering
		\includegraphics[scale=0.1]{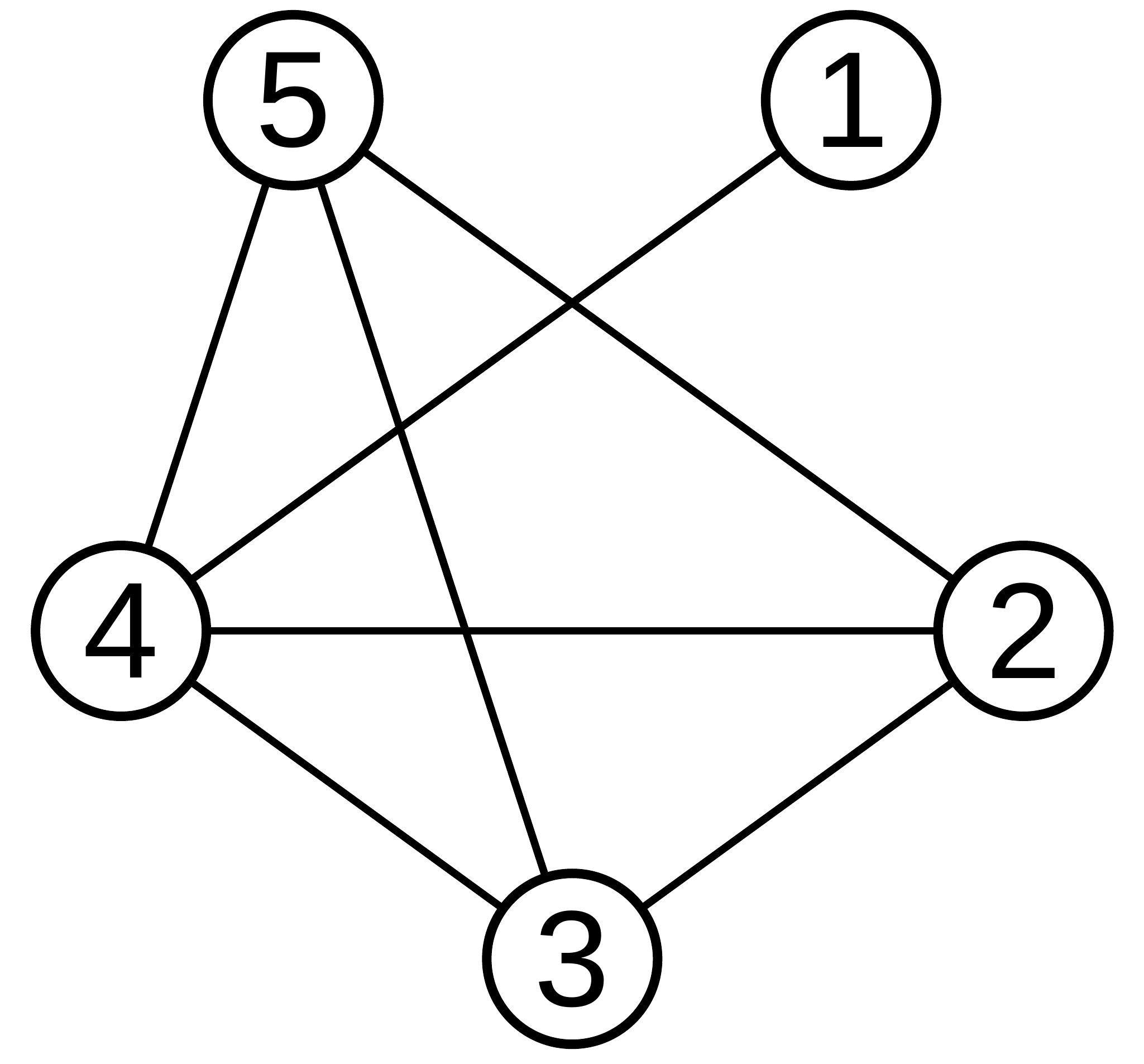}
		\caption{Model 5}
	\end{subfigure}
	\caption{$p=5$}
	\label{fig:p5}
\end{figure}

\begin{figure}[!h]
	\begin{subfigure}{0.32\textwidth}
		\centering
		\includegraphics[scale=0.17]{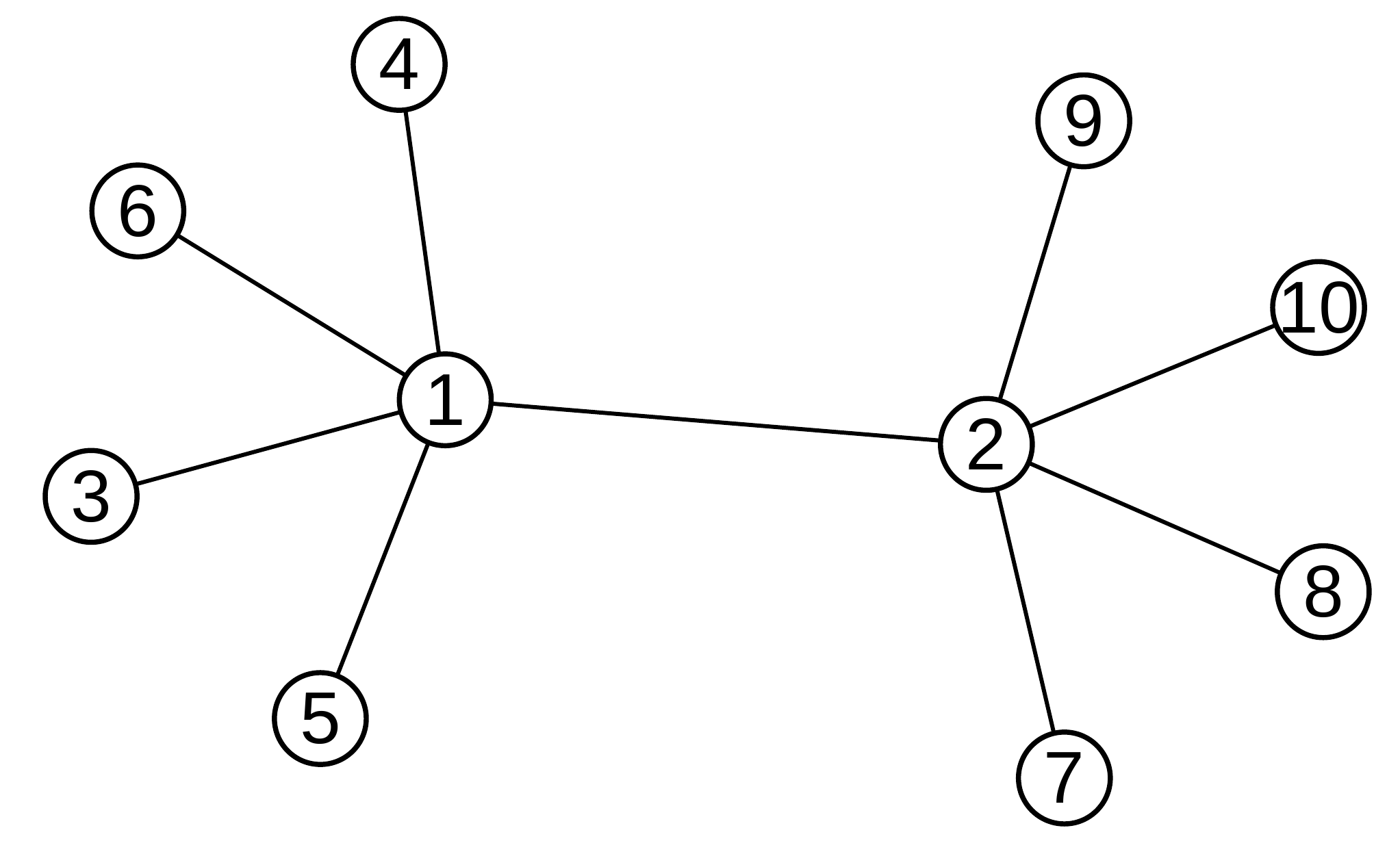}
		\caption{Model 6}
	\end{subfigure} 
	\begin{subfigure}{0.32\textwidth}
		\centering
		\includegraphics[scale=0.17]{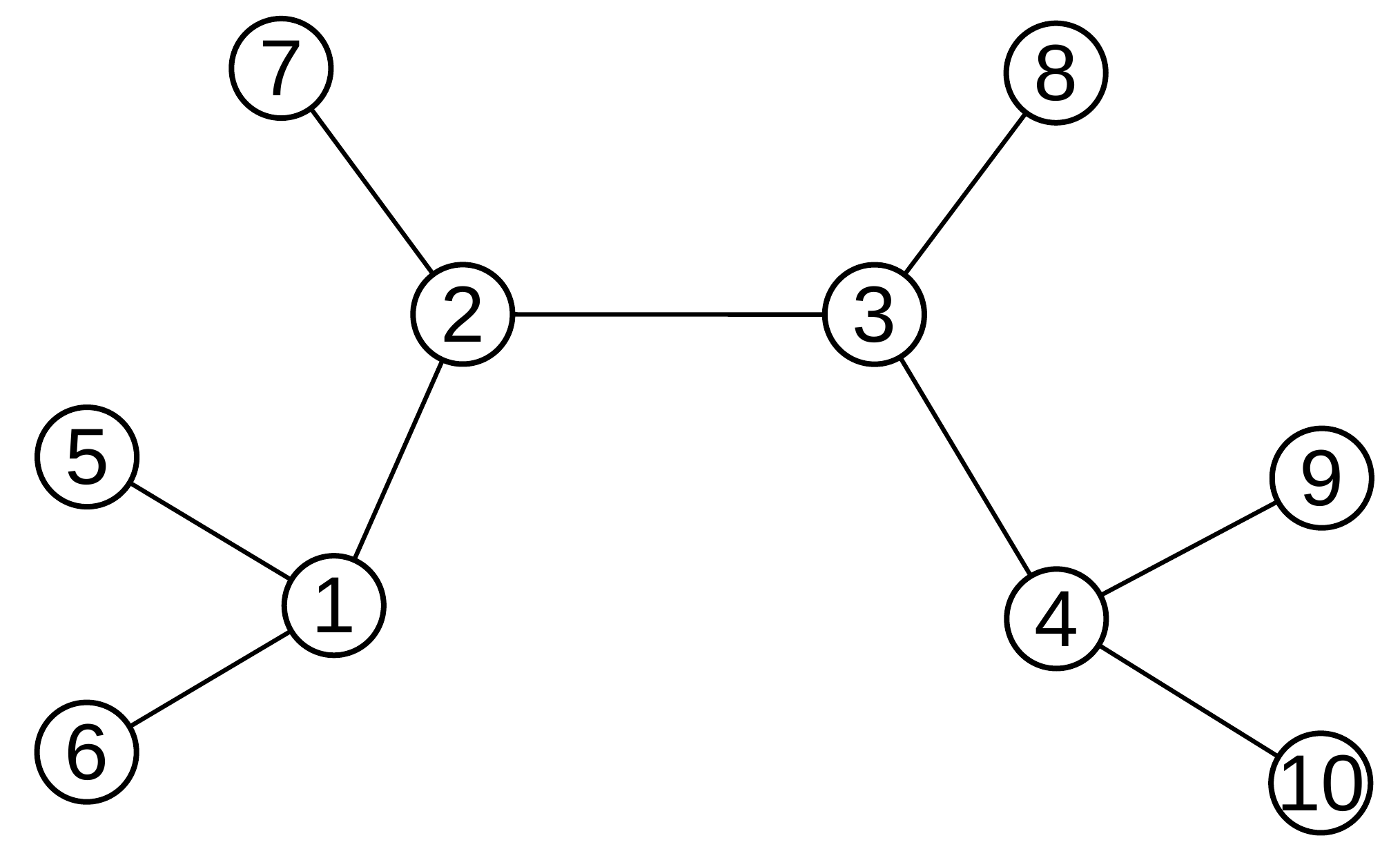}
		\caption{Model 7}
	\end{subfigure}
	\begin{subfigure}{0.32\textwidth}
		\centering
		\includegraphics[scale=0.17]{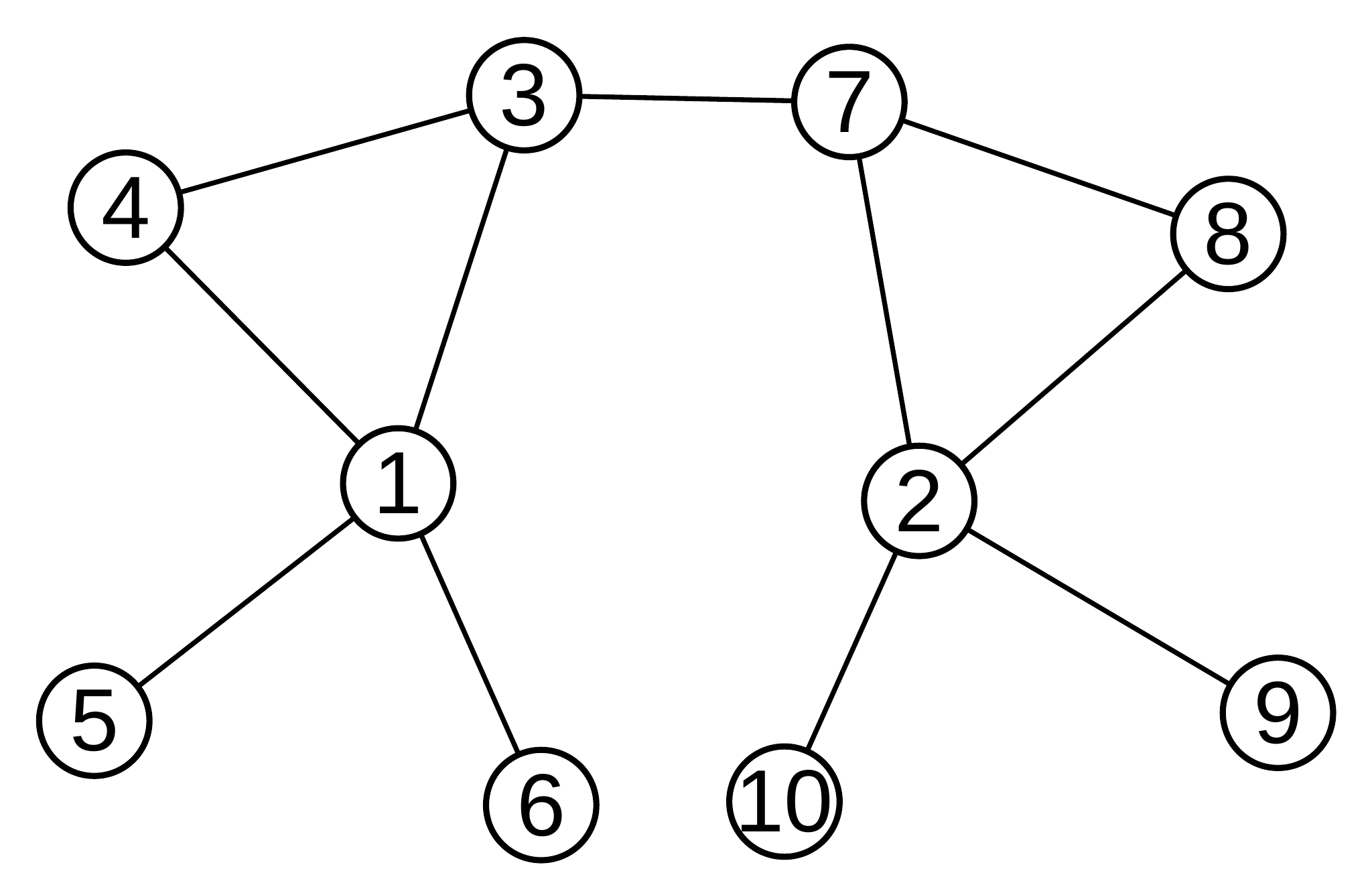}
		\caption{Model 8}
	\end{subfigure}
	
	\hspace{0.15\textwidth}
	\begin{subfigure}{0.3\textwidth}
		\centering
		\includegraphics[scale=0.11]{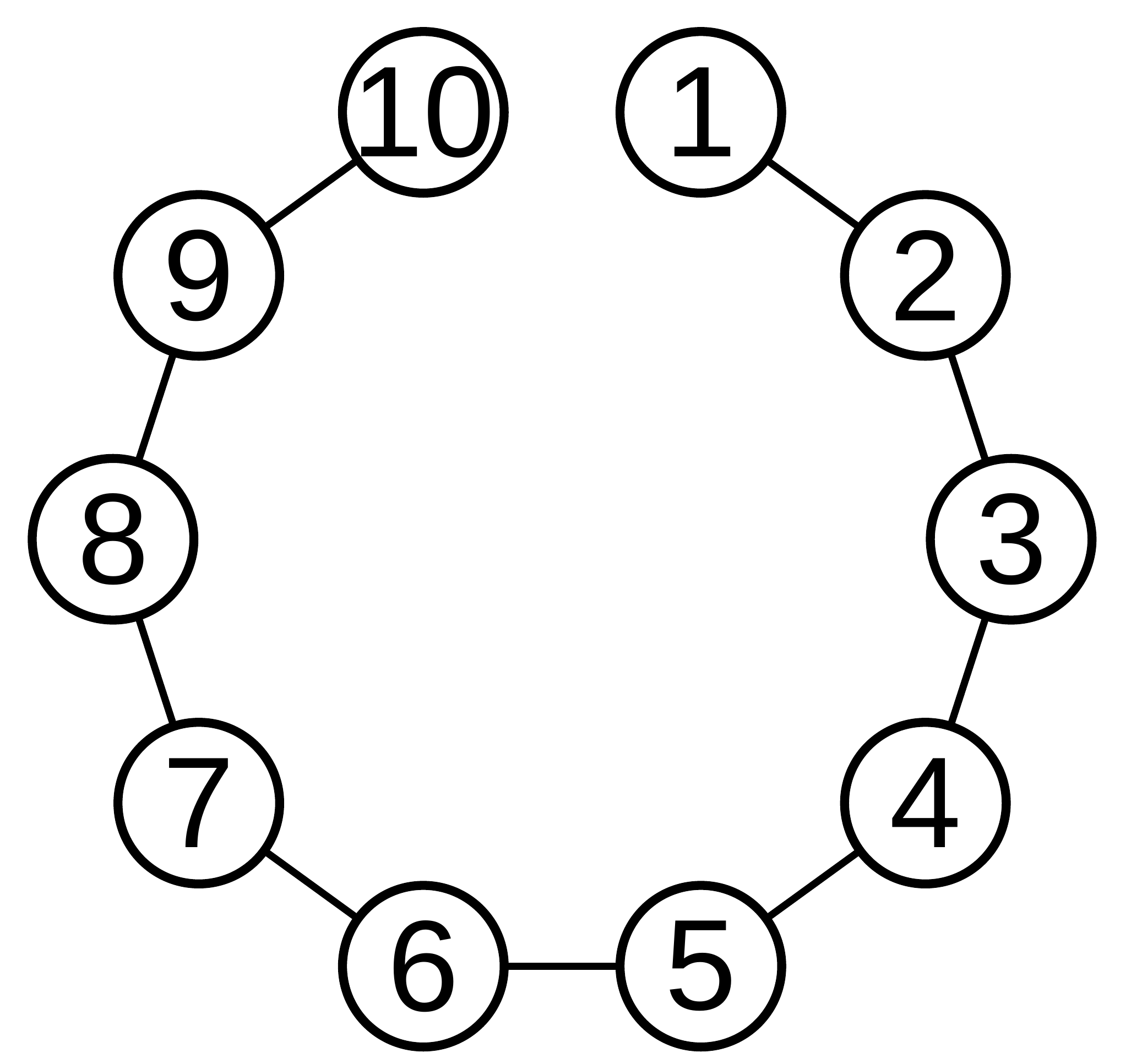}
		\caption{Model 9}
	\end{subfigure}
	\begin{subfigure}{0.32\textwidth}
		\centering
		\includegraphics[scale=0.16]{model10.pdf}
		\caption{Model 10}
	\end{subfigure}
	\caption{$p=10$}
	\label{fig:p10}
\end{figure}

Both estimators involve choosing the $\lambda_T$ tuning parameters. We use a 2-folds cross-validation procedure to tune $\lambda_T$. Specifically, we use the Kullback-Leibler (KL) loss averaged over the two-folds to evaluate predictive accuracies. Equation (\ref{KLcv}) gives the KL loss between the estimated $\hat{\Omega}$ from the training set versus the estimated $\Omega$ from the validation set. 

\begin{align}
KL(\lambda_{T}) = \log \det(\Omega) - \log \det \hat{\Omega} + \tr(\hat{\Omega} \Omega^{-1}) -  p \label{KLcv}
\end{align}

We report the simulation result in Table \ref{cvmc1_T20} for $T=20$ and Table \ref{cvmc1_T50} for $T=50$. In the table, the (a) columns corresponds to SGLASSO whereas the (b) columns refer to GLASSO. In Columns 1(a) and 1(b), we report the optimal $\lambda_T$ as determined by cross-validations, averaged across the 1,000 replications. In Columns 2(a) and 2(b), we report the Kullback-Leibler loss averaged across 1,000 replications. The KL loss between $\hat{\Omega}$ and $\Omega_{0}$ is given by $KL(\hat{\Omega},\Omega_{0}) = \log \det(\Omega_{0}) - \log \det \hat{\Omega} + \tr(\hat{\Omega} \Omega_{0}^{-1}) -  p$. In Columns 3(a) and 3(b), we report the average Frobenius loss between $\hat{\Omega}$ and $\Omega_{0}$.

In the last two columns of Tables \ref{cvmc1_T20} and \ref{cvmc1_T50}, we report the accuracy of graph recovery using SGLASSO and GLASSO. The Kullback-Leibler loss and the Frobenius norms may not fully capture how accurately the zeros are recovered. We introduce an additional metric: $F_{1}$ score is  $F_{1} = \frac{2 \text{precision}\cdot \text{recall}}{\text{precision}+\text{recall}}$, where {\em precision} is the ratio of true positives (TP) to all predicted positives (TP + FP), {\em recall} is the ratio of true positives to all actual positives (TP + FN). The notations are further explained in the confusion matrix in Table \ref{confusion}. Alternatively, the $F_{1}$ score can be written as $F_{1} = \frac{2TP}{2TP + FP + FN}$. 

\begin{table}[h!]
	\centering
	\begin{tabular}{ccc}
		\toprule
		& Predicted value is 1 & Predicted value is 0 \\
		\hline
		Actual value is 1 & True positive (TP) & False negative (FN) \\
		Actual value is 0 & False positive (FP) & True negative (TN) \\
		\bottomrule
	\end{tabular}
	\caption{Confusion matrix} \label{confusion}
\end{table}

Hence, the $F_{1}$ score measures the quality of a binary classifier by equally balancing both the precision and the recall of a classifier. The larger the $F_{1}$ score is, the better the classifier is. The $F_{1}$ score is commonly used in machine learning to evaluate binary classifiers. For instance, the Yelp competition uses the $F_1$ score as a metric to rank competing models.\footnote{https://www.kaggle.com/c/yelp-restaurant-photo-classification} The $F_{1}$ score is favored over the metric $Accuracy = \frac{TP+TN}{TP + TN + FP + FN}$ especially in our current setting where the graphical models are sparse. This is because a model that naively predicts all negatives will obtain a high {\em Accuracy} score just because there are many actual negatives, and the TN term dominates the {\em Accuracy} score.
%

The conclusion from Tables \ref{cvmc1_T20} and  \ref{cvmc1_T50}   is that SGLASSO achieves significantly lower KL and Frobenius losses across all models. Moreover in in terms of graph accuracy, SGLASSO also outperforms GLASSO, as indicated by the $F_{1}$ scores. We  also observe that as we increase $T=20$ to $T=50$, the optimal tuning parameter $\lambda$ decreases, the KL and Frobenius losses also decrease, while the $F_1$ score increases. This makes sense because the likelihood becomes more informative as the sample size increases, and therefore the need for regularization decreases.

\subsection{Minimum Kullback-Leibler losses}

To abstract away from the effects of cross-validations, we consider the lowest Kullback-Leibler losses that can be achieved by our estimator versus the GLASSO. Specifically, we vary $\lambda_{T}$ and at each $\lambda_{T}$, we compute the KL and Frobenius losses between the true $\Omega_{0}$ and $\hat{\Omega}$. We see from Table \ref{kltable} that the superior performance of SGLASSO over GLASSO exists after taking away the randomness due to cross-validations. The lowest possible KL losses achievable by our estimator appears to be smaller, than the corresponding KL losses for the GLASSO.

\subsubsection{Recovering high-degree sparsities}



Theorem \ref{aequi} says that the SGLASSO prioritizes recovering the sparsity between nodes that have higher degrees. We now show some numerical evidence that lends support to this. For each model, we examined the pair of nodes $(i,j)$ such that $d_{i} + d_{j}$ is largest and $\Omega_{0,ij} = 0$, where $d_{i} = \sum_{k=1}^{p}|\Omega_{0,ik}|$ is the true weighted degree of node $i$. For instance, for model 1, the $\argmax_{i,j} (d_{i}+d_{j})$ such that $\Omega_{0,ij} =0$ would correspond to the pairs of nodes $(1,4)$ and $(2,5)$. For simplicity, if there are multiple pairs of nodes that maximize $d_{i}+d_{j}$, we will pick the pair of nodes that comes first when the adjacency matrix is vectorized. 

In each of the model (set $T=20$), we calculate the fraction of times that GLASSO and SGLASSO correctly recover $\Omega_{0,ij} =0$ for the largest $d_{i} + d_{j}$. We plot this result in Figure \ref{fig:prob}, which shows that at any given $\lambda$, the probability of correctly recovering $\Omega_{0,ij} =0$ for high-degree nodes is greater when SGLASSO is used, compared to GLASSO.

\begin{table}[!h]
	{\small
		\makebox[\textwidth][c]{
			\begin{tabu}{cccccccccccc}
				\toprule
				Model & \multicolumn{2}{c}{Optimal $\lambda_{T}$} && \multicolumn{2}{c}{KL}  & & \multicolumn{2}{c}{Frobenius} & & \multicolumn{2}{c}{$F_{1}$ score}  \\
				\cmidrule{2-3} \cmidrule{5-6} \cmidrule{8-9} \cmidrule{11-12}
				& (a)  & (b)  && (a) & (b) && (a) & (b)  && (a) & (b)  \\
				& {\scriptsize SGLASSO}  & {\scriptsize GLASSO}  && {\scriptsize SGLASSO} &  {\scriptsize GLASSO} && {\scriptsize SGLASSO} &  {\scriptsize GLASSO}  && {\scriptsize SGLASSO} &  {\scriptsize GLASSO} \\
				\hline
				(1) & 0.181 & 0.342 && 0.430 & 0.483  && 0.896 & 0.948 && 0.439 & 0.367  \\
				\rowfont{\footnotesize} 
				& (0.118) & (0.133) && (0.207) & (0.215) && (0.223) & (0.214) && (0.248) & (0.259)  \\ 
				(2) & 0.180 & 0.341 && 0.419 & 0.469  && 0.858 & 0.906 && 0.400 & 0.349  \\ 
				\rowfont{\footnotesize}
				& (0.118) & (0.136) && (0.205) & (0.215) && (0.232) & (0.226) && (0.254) & (0.276)  \\ 
				(3) & 0.183 & 0.345 && 0.438 & 0.492  && 0.904 & 0.956 && 0.415 & 0.353  \\ 
				\rowfont{\footnotesize}
				& (0.119) & (0.138) && (0.210) & (0.206) && (0.216) & (0.204) && (0.240) & (0.262)  \\ 
				(4) & 0.181 & 0.341 && 0.422 & 0.471  && 0.859 & 0.905 && 0.396 & 0.325  \\ 
				\rowfont{\footnotesize}
				& (0.119) & (0.137) && (0.223) & (0.208) && (0.234) & (0.210) && (0.253) & (0.274)  \\ 
				(5) & 0.188 & 0.348 && 0.453 & 0.504  && 0.985 & 1.032 && 0.436 & 0.349  \\ 
				\rowfont{\footnotesize}
				& (0.121) & (0.138) && (0.195) & (0.204) && (0.189) & (0.181) && (0.262) & (0.270)  \\ 
				\hline  
				\\
				(6) & 0.203 & 0.397 && 0.913 & 1.050  && 1.248 & 1.340 && 0.332 & 0.296  \\ 
				\rowfont{\footnotesize}
				& (0.095) & (0.106) && (0.270) & (0.446) && (0.172) & (0.312) && (0.153) & (0.172)  \\ 
				(7) & 0.206 & 0.398 && 0.909 & 1.067  && 1.264 & 1.377 && 0.314 & 0.274  \\ 
				\rowfont{\footnotesize}
				& (0.094) & (0.110) && (0.289) & (0.579) && (0.204) & (0.507) && (0.141) & (0.163)  \\ 
				(8) & 0.210 & 0.402 && 0.939 & 1.089  && 1.327 & 1.431 && 0.308 & 0.261  \\ 
				\rowfont{\footnotesize}
				& (0.098) & (0.110) && (0.287) & (0.555) && (0.192) & (0.431) && (0.145) & (0.160)  \\ 
				(9) & 0.208 & 0.400 && 0.936 & 1.089  && 1.291 & 1.397 && 0.317 & 0.260  \\ 
				\rowfont{\footnotesize}
				& (0.098) & (0.109) && (0.297) & (0.624) && (0.202) & (0.470) && (0.149) & (0.161)  \\ 
				(10) & 0.207 & 0.399 && 0.946 & 1.096  && 1.353 & 1.456 && 0.336 & 0.272  \\ 
				\rowfont{\footnotesize}
				& (0.094) & (0.109) && (0.278) & (0.527) && (0.183) & (0.414) && (0.143) & (0.162)  \\ 
				(11) & 0.198 & 0.389 && 1.201 & 1.390  && 1.569 & 1.701 && 0.403  & 0.341  \\ 
				\rowfont{\footnotesize}
				& (0.097) & (0.117) && (0.312) & (0.723) && (0.186) & (0.522) && (0.143) & (0.164)  \\ 
				\bottomrule
			\end{tabu}
		}
		\caption{Averages and standard errors from 1,000 replications, $T=20$. Columns 1(a) and 1(b) report the optimal $\lambda_T$ as determined from cross-validation for (a) SGLASSO and (b) GLASSO. Columns 2(a) and 2(b) are the average Kullback-Leibler losses. Columns 3(a) and 3(b) report the average Frobenius losses.  The last two columns report the $F_{1}$ scores, which is a metric for how accurately the graphical model is estimated.}\label{cvmc1_T20}}
\end{table}

\begin{table}[!h]
	{\small
		\makebox[\textwidth][c]{
			\begin{tabu}{cccccccccccc}
				\toprule
				Model & \multicolumn{2}{c}{Optimal $\lambda_{T}$} && \multicolumn{2}{c}{KL}  & & \multicolumn{2}{c}{Frobenius} & & \multicolumn{2}{c}{$F_{1}$ score}  \\
				\cmidrule{2-3} \cmidrule{5-6} \cmidrule{8-9} \cmidrule{11-12}
				& (a)  & (b)  && (a) & (b) && (a) & (b)  && (a) & (b)  \\
				& {\scriptsize SGLASSO}  & {\scriptsize GLASSO}  && {\scriptsize SGLASSO} &  {\scriptsize GLASSO} && {\scriptsize SGLASSO} &  {\scriptsize GLASSO}  && {\scriptsize SGLASSO} &  {\scriptsize GLASSO} \\
				\hline
				(1) & 0.088 & 0.179 && 0.246 & 0.265  && 0.710 & 0.736 && 0.564 & 0.541  \\ 
				\rowfont{\footnotesize} 
				& (0.054) & (0.076) && (0.084) & (0.091) && (0.129) & (0.134) && (0.225) & (0.248)  \\ 
				(2) & 0.090 & 0.183 && 0.232 & 0.252  && 0.663 & 0.692 && 0.548 & 0.517  \\ 
				\rowfont{\footnotesize} 
				& (0.052) & (0.075) && (0.083) & (0.090) && (0.127) & (0.130) && (0.233) & (0.248)  \\ 
				(3) & 0.090 & 0.181 && 0.249 & 0.269  && 0.715 & 0.744 && 0.565 & 0.530  \\ 
				\rowfont{\footnotesize} 
				& (0.053) & (0.077) && (0.083) & (0.091) && (0.124) & (0.131) && (0.215) & (0.239)  \\ 
				(4) & 0.090 & 0.183 && 0.234 & 0.256  && 0.666 & 0.696 && 0.540 & 0.504  \\ 
				\rowfont{\footnotesize} 
				& (0.053) & (0.076) && (0.086) & (0.095) && (0.127) & (0.135) && (0.222) & (0.250)  \\ 
				(5) & 0.089 & 0.181 && 0.268 & 0.290  && 0.800 & 0.831 && 0.568 & 0.521  \\ 
				\rowfont{\footnotesize} 
				& (0.054) & (0.079) && (0.081) & (0.092) && (0.125) & (0.138) && (0.228) & (0.252)  \\ 
				\hline  
				\\
				(6) & 0.106 & 0.228 && 0.557 & 0.615  && 1.024 & 1.076 && 0.485 & 0.479  \\ 
				\rowfont{\footnotesize}
				& (0.038) & (0.050) && (0.118) & (0.132) && (0.109) & (0.116) && (0.135) & (0.160)  \\ 
				(7) & 0.109 & 0.230 && 0.551 & 0.607  && 1.030 & 1.084 && 0.453 & 0.437  \\ 
				\rowfont{\footnotesize}
				& (0.040) & (0.048) && (0.109) & (0.119) && (0.109) & (0.111) && (0.145) & (0.156)  \\ 
				(8) & 0.110 & 0.231 && 0.583 & 0.642  && 1.105 & 1.160 && 0.457 & 0.419  \\ 
				\rowfont{\footnotesize}
				& (0.040) & (0.050) && (0.114) & (0.129) && (0.112) & (0.117) && (0.128) & (0.147)  \\ 
				(9) & 0.109 & 0.230 && 0.577 & 0.635  && 1.061 & 1.117 && 0.460 & 0.438  \\ 
				\rowfont{\footnotesize}
				& (0.040) & (0.050) && (0.106) & (0.118) && (0.111) & (0.114) && (0.137) & (0.158)  \\ 
				(10) & 0.109 & 0.230 && 0.593 & 0.651  && 1.137 & 1.190 && 0.466 & 0.437  \\ 
				\rowfont{\footnotesize}
				& (0.040) & (0.052) && (0.115) & (0.128) && (0.113) & (0.117) && (0.136) & (0.159)  \\ 
				(11) & 0.091 & 0.206  && 0.747 & 0.807  && 1.312 & 1.370 && 0.563 & 0.541  \\ 
				\rowfont{\footnotesize}
				& (0.041) & (0.053) && (0.153) & (0.167) && (0.128) & (0.133) && (0.120) & (0.137)  \\ 
				\bottomrule
			\end{tabu}
		}
		\caption{Averages and standard errors from 1,000 replications, $T=50$. Columns 1(a) and 1(b) report the optimal $\lambda_T$ as determined from cross-validation for (a) SGLASSO and (b) GLASSO. Columns 2(a) and 2(b) are the average Kullback-Leibler losses. Columns 3(a) and 3(b) report the average Frobenius losses.  The last two columns report the $F_{1}$ scores, which is a metric for how accurately the graphical model is estimated.}\label{cvmc1_T50}}
\end{table}

\begin{table}[h!]
	{\small
		\centering
		\begin{tabu}{cccccccc}
			\toprule
			Model & \multicolumn{2}{c}{Minimum}  & & \multicolumn{2}{c}{Minimum} & & \multicolumn{1}{c}{Percentage}  \\
			& \multicolumn{2}{c}{KL}  & & \multicolumn{2}{c}{Frobenius} & & \multicolumn{1}{c}{Dominance}  \\
			\cmidrule{2-3} \cmidrule{5-6} 
			& (a)  & (b)  && (a) & (b) &&  \\
			& {\footnotesize SGLASSO}  & {\footnotesize GLASSO}  && {\footnotesize SGLASSO} &  {\footnotesize GLASSO} &&  \\
			\hline
			(1) & 0.340 & 0.376 && 0.779 & 0.821  && 0.99 \\ 
			& (0.106) & (0.120) && (0.096) & (0.116)  & \\ 
			(2) & 0.324 & 0.361 && 0.734 & 0.780  && 0.99 \\ 
			& (0.107) & (0.120) && (0.102) & (0.121)  & \\ 
			(3) & 0.341 & 0.378 && 0.778 & 0.821  && 0.98 \\ 
			& (0.105) & (0.119) && (0.095) & (0.115)  & \\ 
			(4) & 0.326 & 0.362 && 0.734 & 0.779  && 0.99 \\ 
			& (0.106) & (0.120) && (0.102) & (0.121)  & \\ 
			(5) & 0.362 & 0.399 && 0.873 & 0.908  && 0.98 \\ 
			& (0.104) & (0.118) && (0.081) & (0.099)  & \\ 
			(6) & 0.789 & 0.846 && 1.134 & 1.177  && 0.99 \\ 
			& (0.162) & (0.172) && (0.092) & (0.103)  & \\ 
			(7) & 0.771 & 0.831 && 1.136 & 1.182  && 1.00 \\ 
			& (0.159) & (0.168) && (0.091) & (0.102)  & \\ 
			(8) & 0.805 & 0.865 && 1.199 & 1.242  && 1.00 \\ 
			& (0.158) & (0.168) && (0.087) & (0.097)  & \\ 
			(9) & 0.800 & 0.860 && 1.166 & 1.211  && 1.00 \\ 
			& (0.161) & (0.172) && (0.091) & (0.100)  & \\ 
			(10) & 0.824 & 0.887 && 1.234 & 1.279  && 0.99 \\ 
			& (0.162) & (0.176) && (0.085) & (0.101)  & \\ 
			(11) & 1.041 & 1.102 && 1.415 & 1.466  && 0.95 \\ 
			& (0.191) & (0.205) && (0.087) & (0.101)  & \\ 
			\bottomrule
		\end{tabu}
		\caption{$T=20$. Minimum possible KL and Frobenius losses across all $\lambda_T$. The numbers reported are the average KL and Frobenius losses evaluated at the $\lambda_T$ that result in the lowest possible values.  Standard deviations across 1,000 replications are reported in parentheses. In the last column, we report the fraction of times in which SGLASSO performed better in terms of KL loss at the corresponding $\lambda_T$.}
		\label{kltable}
	}
\end{table}

\begin{table}[h!]
	{\small
		\centering
		\begin{tabu}{cccccccc}
			\toprule
			Model & \multicolumn{2}{c}{Minimum}  & & \multicolumn{2}{c}{Minimum} & & \multicolumn{1}{c}{Percentage}  \\
			& \multicolumn{2}{c}{KL}  & & \multicolumn{2}{c}{Frobenius} & & \multicolumn{1}{c}{Dominance}  \\
			\cmidrule{2-3} \cmidrule{5-6} 
			& (a)  & (b)  && (a) & (b) &&  \\
			& {\footnotesize SGLASSO}  & {\footnotesize GLASSO}  && {\footnotesize SGLASSO} &  {\footnotesize GLASSO} &&  \\
			\hline
			(1) & 0.199 & 0.211 && 0.606 & 0.634  && 0.92 \\ 
			& (0.063) & (0.065) && (0.095) & (0.098)  & \\ 
			(2) & 0.185 & 0.198 && 0.575 & 0.596  && 0.87 \\ 
			& (0.059) & (0.066) && (0.088) & (0.102)  & \\ 
			(3) & 0.198 & 0.210 && 0.606 & 0.632  && 0.93 \\ 
			& (0.062) & (0.064) && (0.093) & (0.096)  & \\ 
			(4) & 0.184 & 0.197 && 0.575 & 0.596  && 0.85 \\ 
			& (0.058) & (0.064) && (0.086) & (0.100)  & \\ 
			(5) & 0.218 & 0.231 && 0.679 & 0.697  && 0.94 \\ 
			& (0.063) & (0.069) && (0.086) & (0.098)  & \\ 
			(6) & 0.468 & 0.488 && 0.916 & 0.931  && 0.90 \\ 
			& (0.091) & (0.100) && (0.078) & (0.090)  & \\ 
			(7) & 0.465 & 0.491 && 0.917 & 0.940  && 0.99 \\ 
			& (0.091) & (0.098) && (0.081) & (0.089)  & \\ 
			(8) & 0.497 & 0.522 && 0.987 & 1.005  && 0.96 \\ 
			& (0.085) & (0.093) && (0.073) & (0.083)  & \\ 
			(9) & 0.485 & 0.512 && 0.940 & 0.964  && 0.99 \\ 
			& (0.091) & (0.097) && (0.080) & (0.085)  & \\ 
			(10) & 0.509 & 0.533 && 1.021 & 1.035  && 0.96 \\ 
			& (0.087) & (0.095) && (0.072) & (0.082)  & \\ 
			(11) & 0.652 & 0.675 && 1.203 & 1.204  && 0.86 \\ 
			& (0.104) & (0.114) && (0.067) & (0.082)  & \\ 
			\bottomrule
		\end{tabu}
		\caption{$T=50$. Minimum possible KL and Frobenius losses across all $\lambda_T$. The numbers reported are the average KL and Frobenius losses evaluated at the $\lambda_T$ that result in the lowest possible values.  Standard deviations across 1,000 replications are reported in parentheses. In the last column, we report the fraction of times  in which SGLASSO performed better in terms of KL loss at the corresponding $\lambda_T$.}
		\label{kltable}
	}
\end{table}

\begin{figure}[!h]
	\begin{subfigure}{0.2\textwidth}
		\centering
		\includegraphics[scale=0.3]{star1.pdf}
		\caption{Model 1}
	\end{subfigure} 
	\begin{subfigure}{0.2\textwidth}
		\includegraphics[scale=0.3]{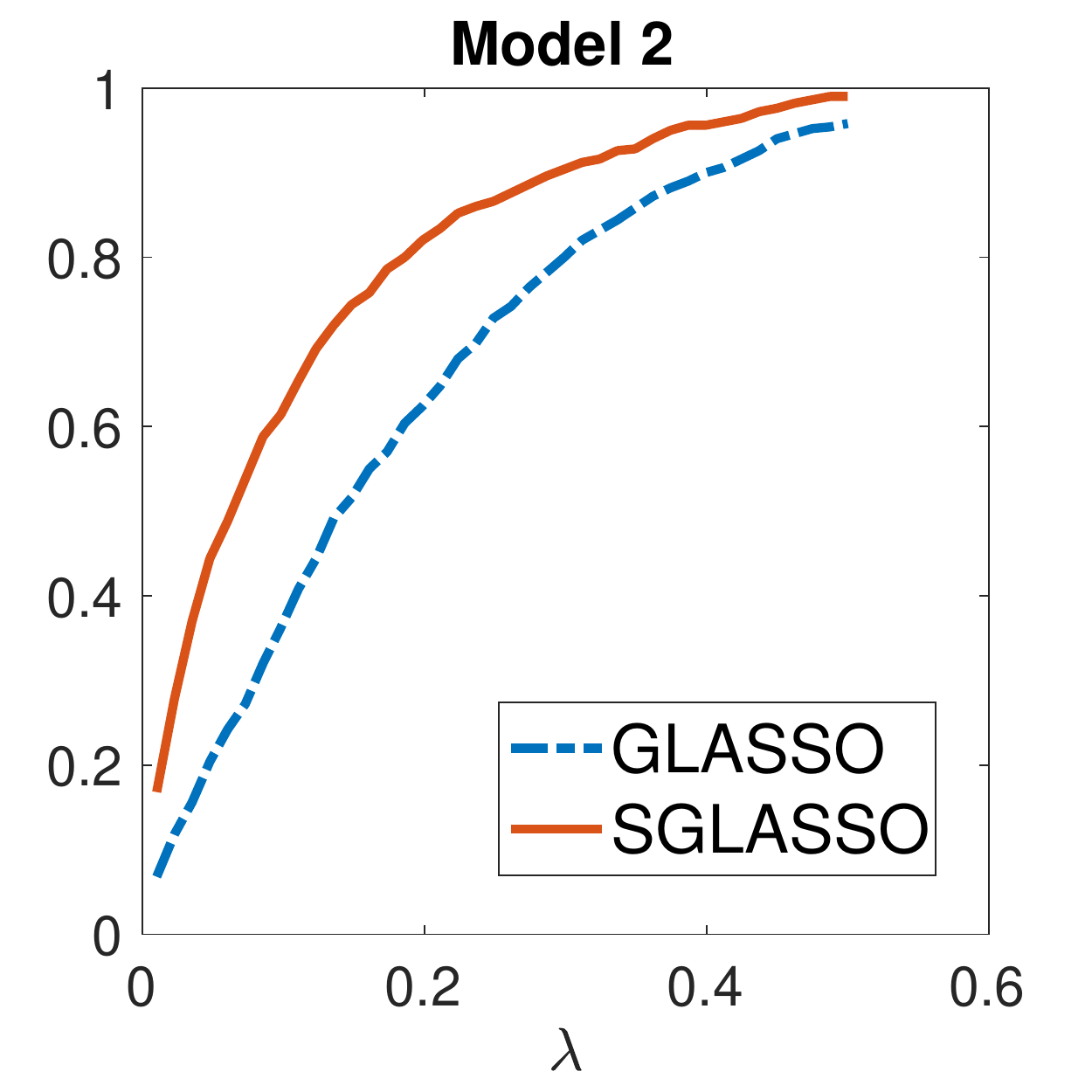}
		\caption{Model 2}
	\end{subfigure}
	\begin{subfigure}{0.2\textwidth}
		\includegraphics[scale=0.3]{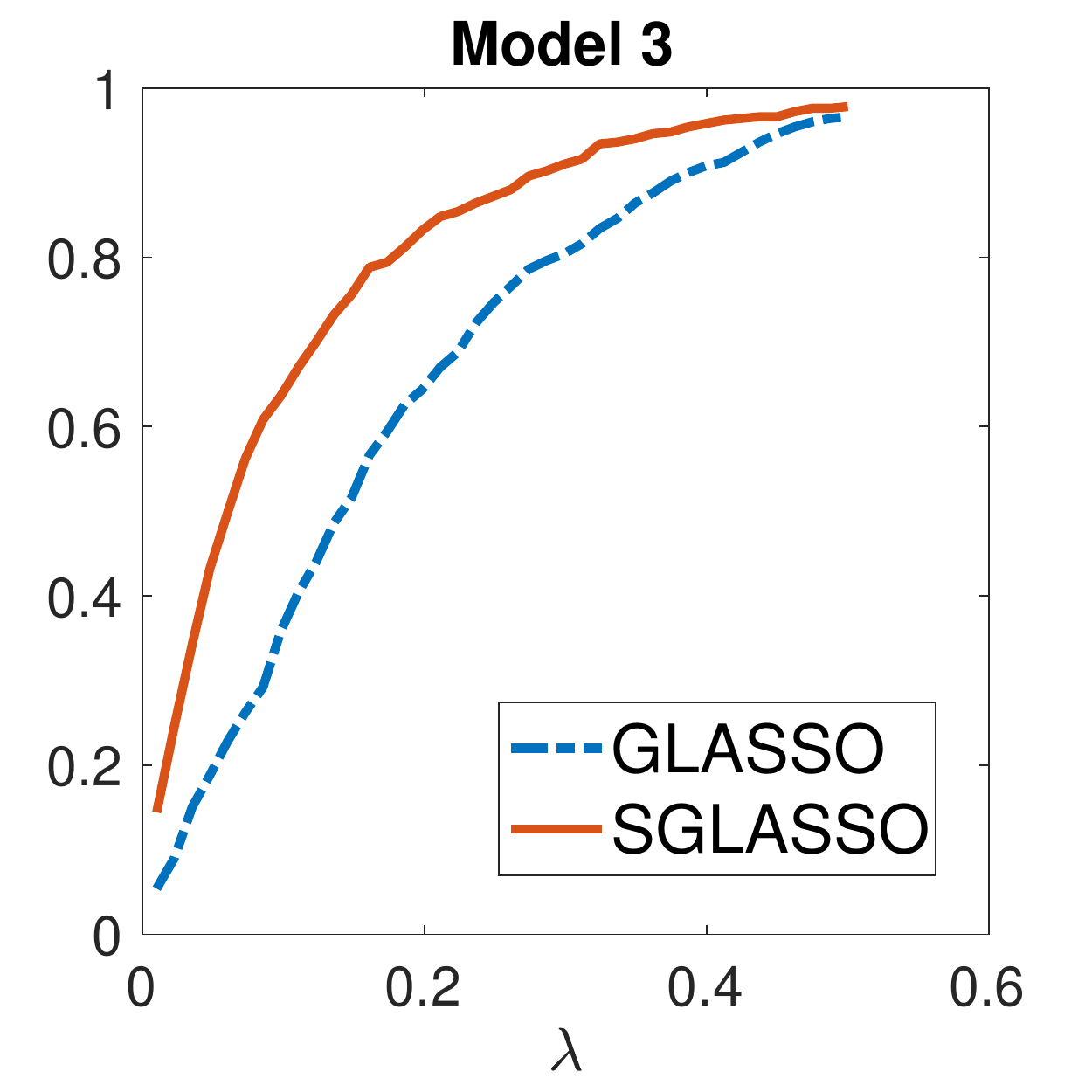}
		\caption{Model 3}
	\end{subfigure}
	\begin{subfigure}{0.2\textwidth}
		\includegraphics[scale=0.3]{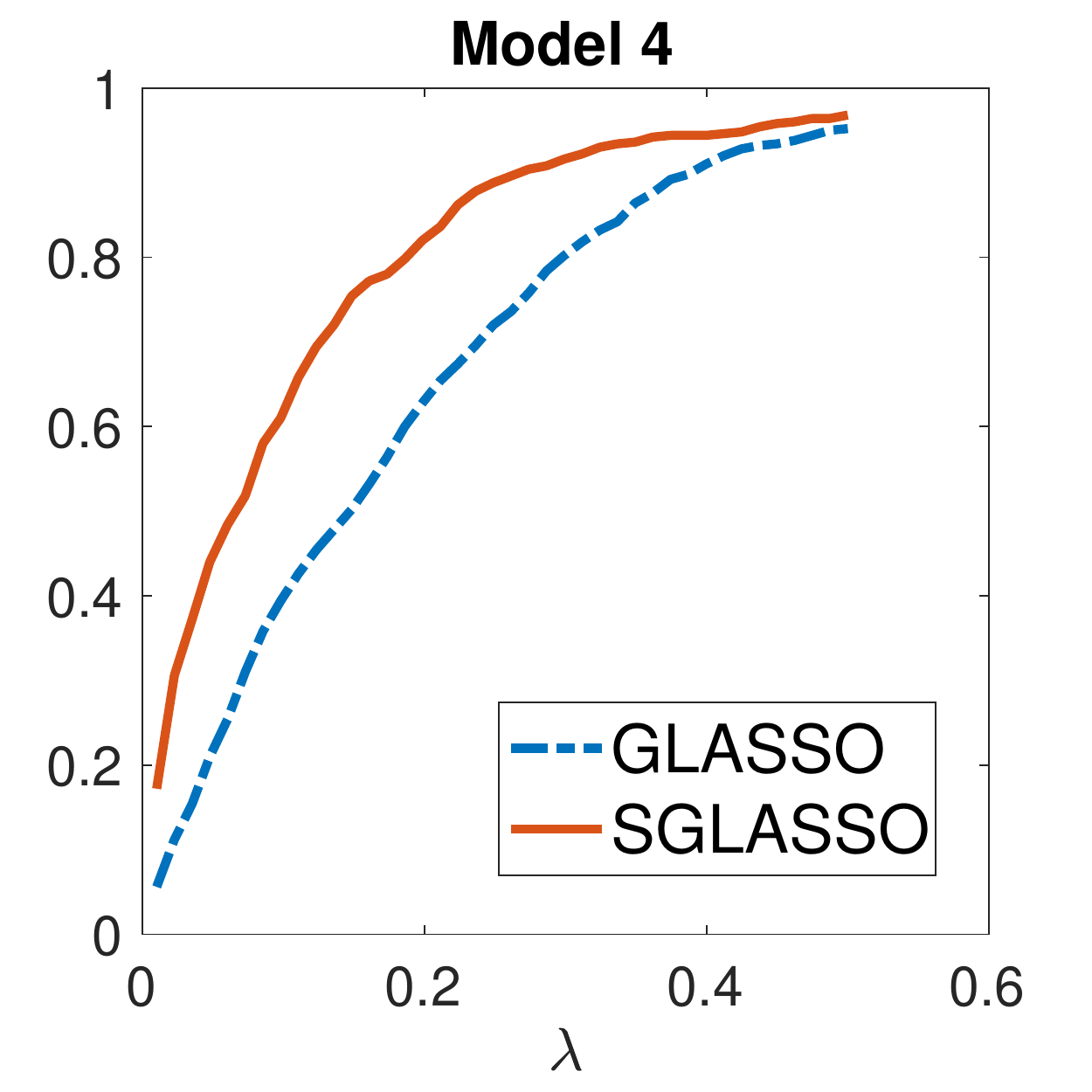}
		\caption{Model 4}
	\end{subfigure} 
	\begin{subfigure}{0.2\textwidth}
		\includegraphics[scale=0.3]{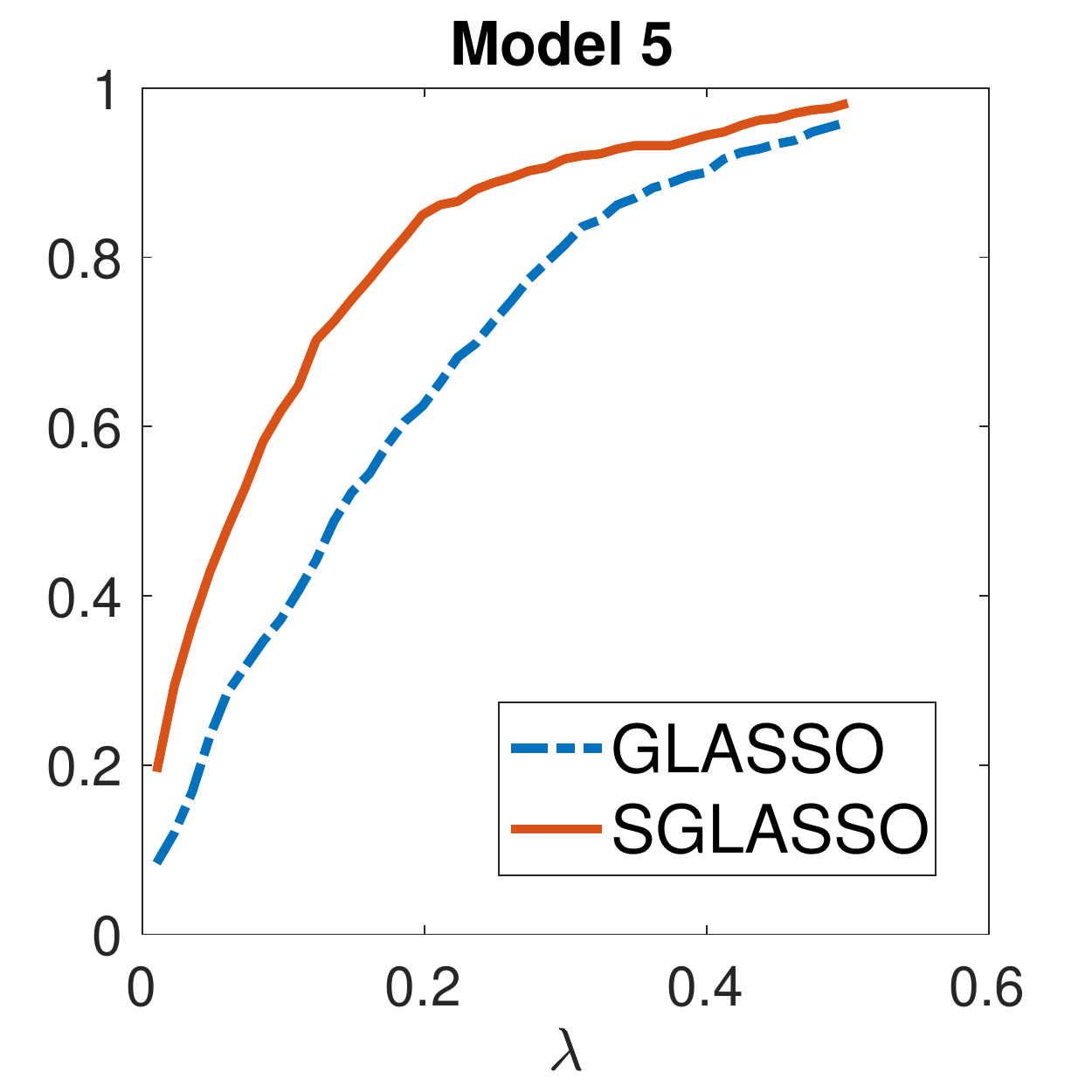}
		\caption{Model 5}
	\end{subfigure}
	\begin{subfigure}{0.2\textwidth}
		\includegraphics[scale=0.3]{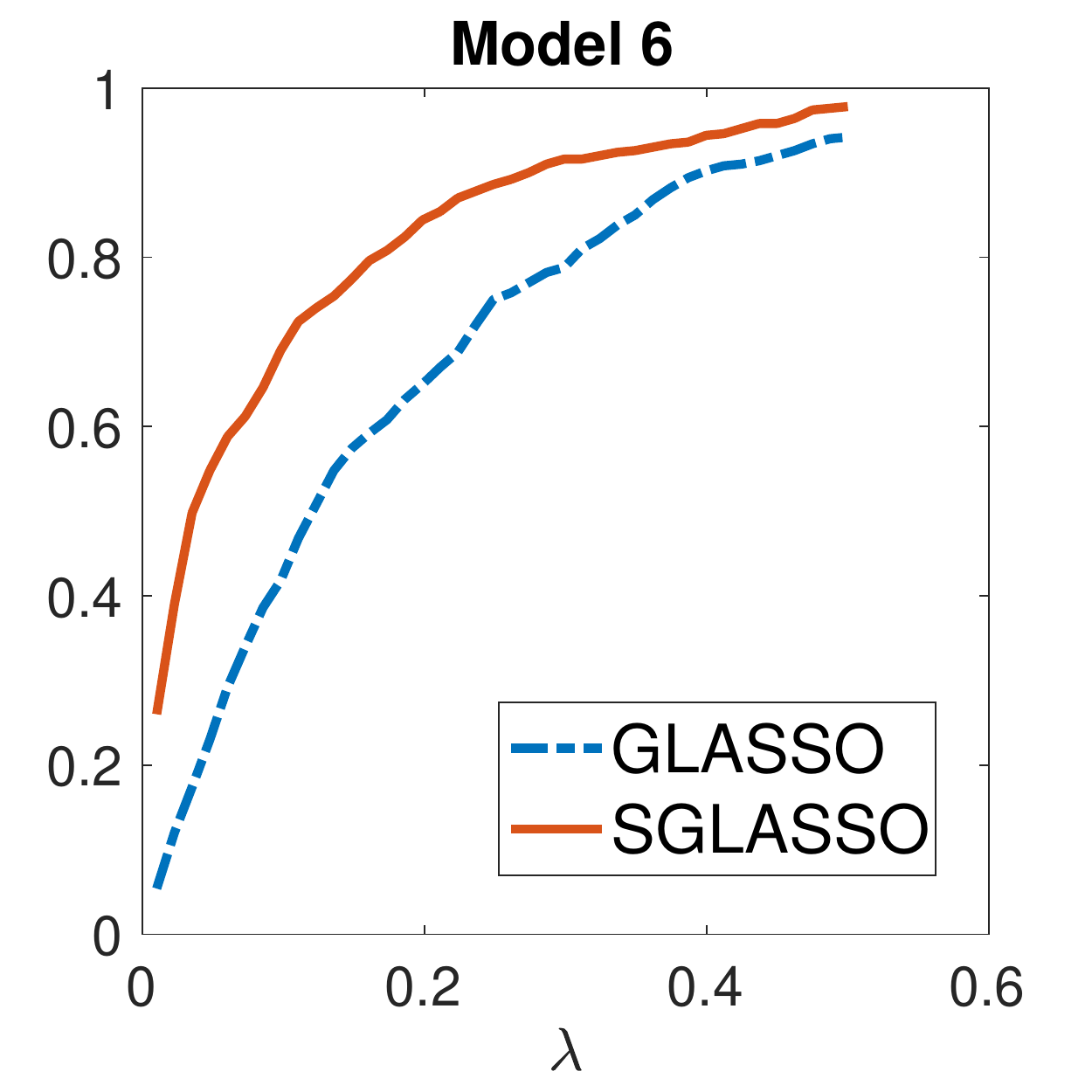}
		\caption{Model 6}
	\end{subfigure}
	\begin{subfigure}{0.2\textwidth}
		\includegraphics[scale=0.3]{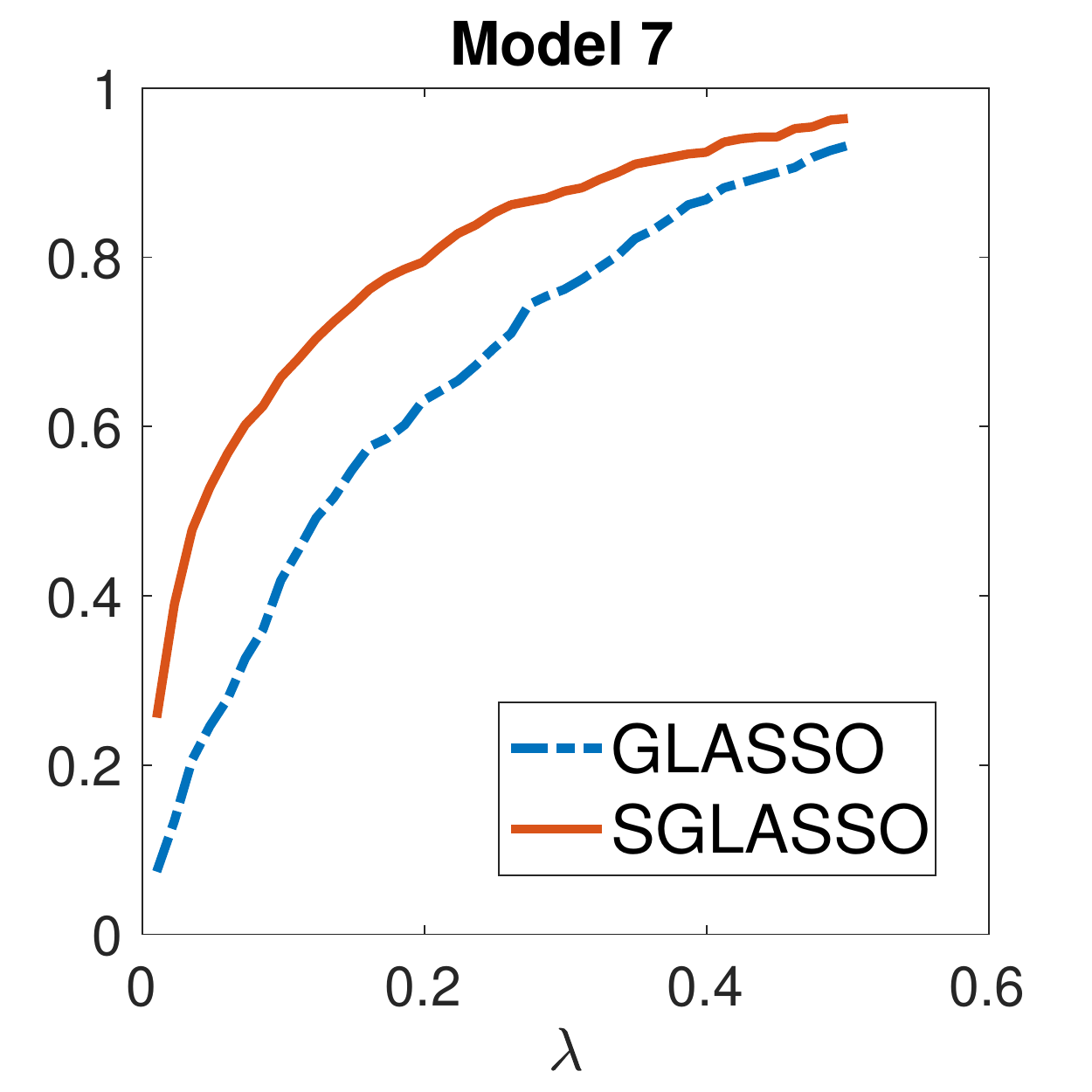}
		\caption{Model 7}
	\end{subfigure} 
	\begin{subfigure}{0.2\textwidth}
		\includegraphics[scale=0.3]{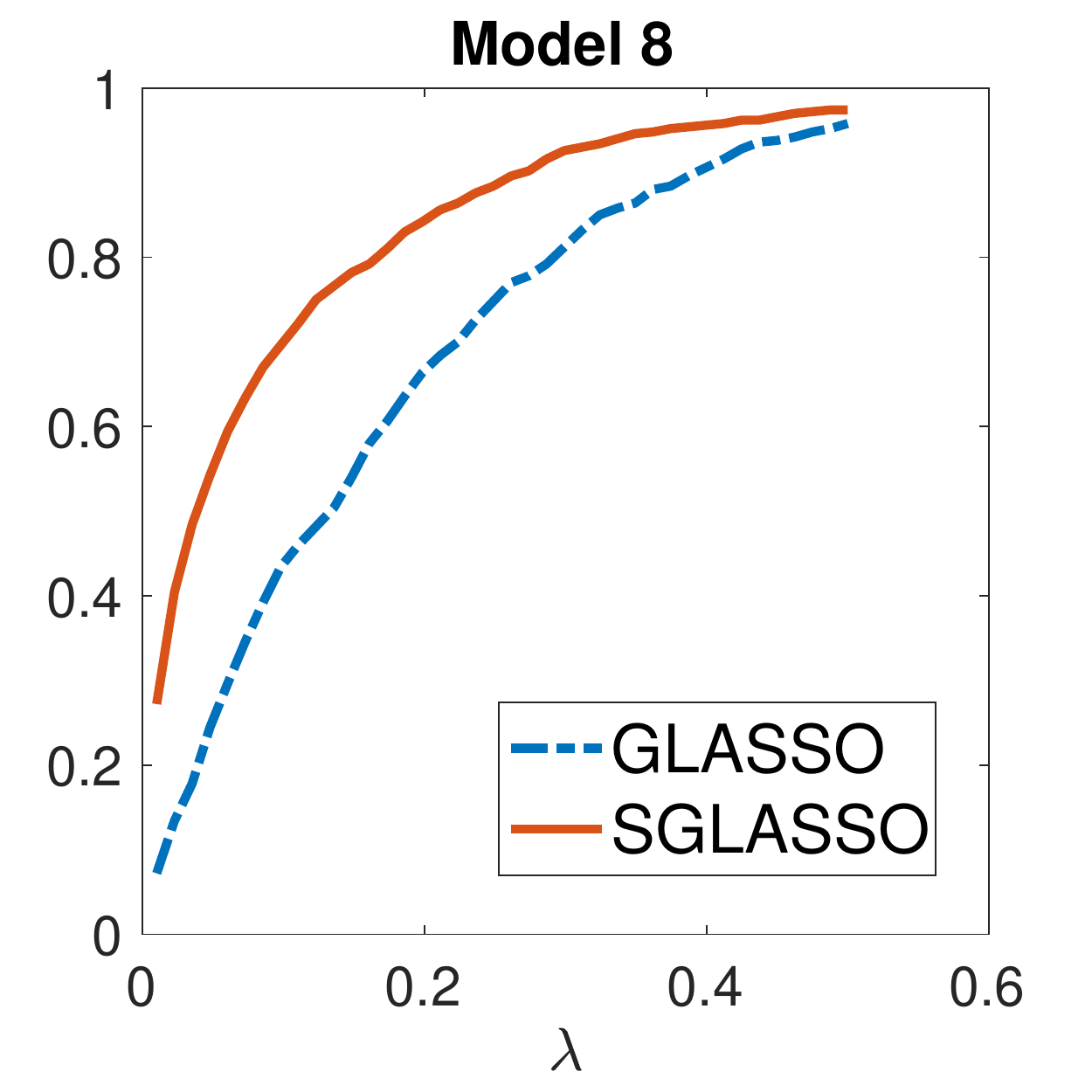}
		\caption{Model 8}
	\end{subfigure}
	\begin{subfigure}{0.3\textwidth}
		\includegraphics[scale=0.3]{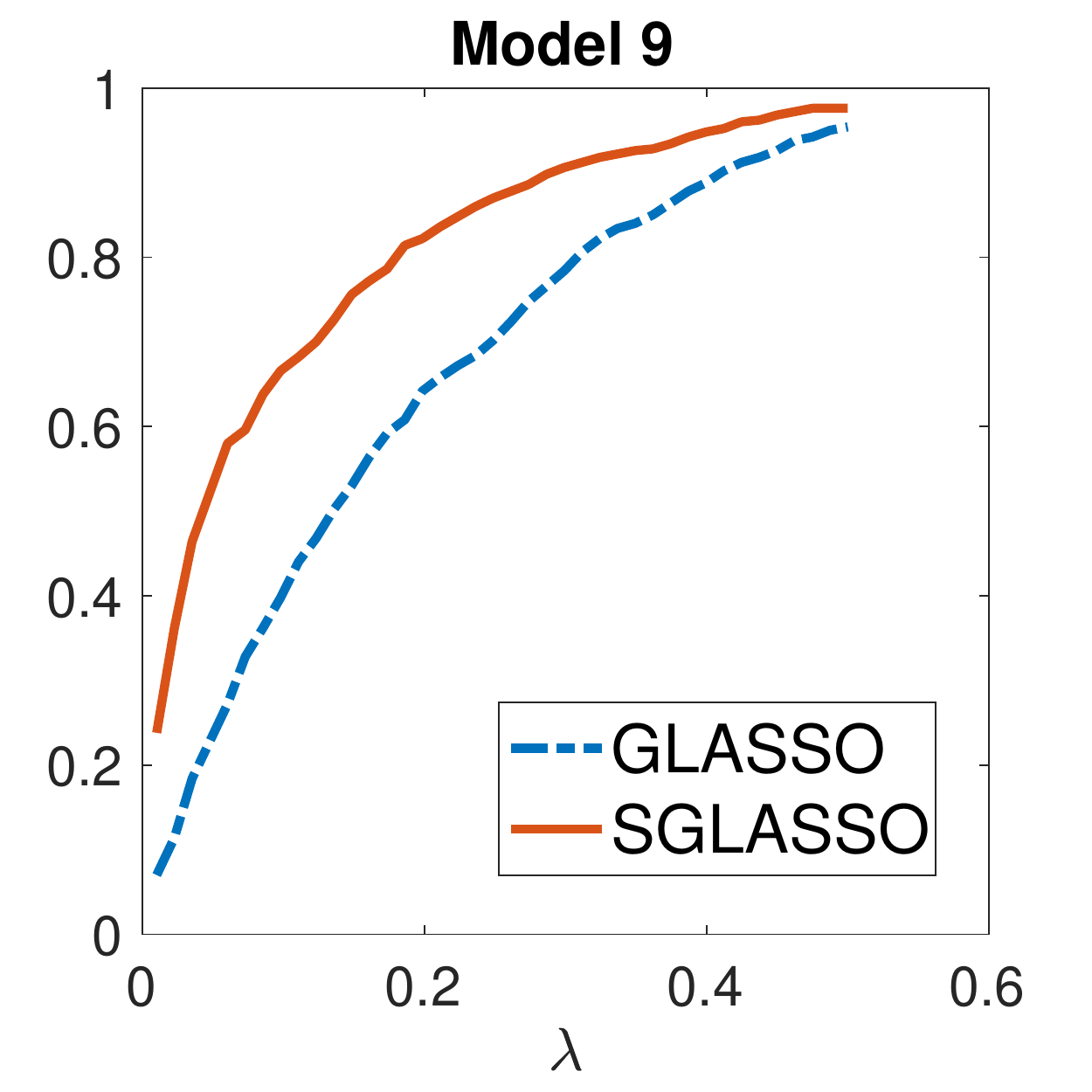}
		\caption{Model 9}
	\end{subfigure}
	\begin{subfigure}{0.3\textwidth}
		\includegraphics[scale=0.3]{star10.pdf}
		\caption{Model 10}
	\end{subfigure}
	\begin{subfigure}{0.3\textwidth}
		\includegraphics[scale=0.3]{star11.pdf}
		\caption{Model 11}
	\end{subfigure}
	\caption{Each figure shows the estimated probability of recovering the non-link $(i,j)$ such that $(d_{i} + d_{j})$ is the largest in each model, as a function of the tuning parameter $\lambda_T$. More precisely, for a given $\lambda_T$, we calculate the number of times (out of the 1,000 replications) that SGLASSO and GLASSO successfully estimate the entry $\Omega_{ij}$ as zero, where $(i,j)$ is such that $\Omega_{0i,j} = 0$ and that $d_{i} + d_{j}$ has the highest value among all non-links.}\label{fig:prob}
\end{figure}

}

\FloatBarrier


\end{document}